\pdfoutput=1
\documentclass[final]{siamltex}
\usepackage{pdfpages}
\usepackage{amssymb}
\usepackage{empheq}
\usepackage{cases}
\usepackage{amsmath}
\usepackage{caption,lipsum}
\usepackage{stmaryrd}
\usepackage{graphicx,epsfig}
\usepackage{epstopdf}
\usepackage{tabularx}
\usepackage{color}
\usepackage{empheq,float}
\usepackage{tikz}
\usetikzlibrary{matrix}
\DeclareGraphicsExtensions{-eps-converted-to.pdf}
\input psfig.sty
\def\RR{{\mathbb R}}
\def\NN{{\mathbb N}}

\newcommand{\bx}{\bf x}
\newcommand{\fe}{\mathrm{e}}
\newcommand{\ud}{\mathrm{d}}
\newcommand{\re}{\mathrm{Re}}

\def\eps{\varepsilon}

\def\pa{\partial}

\def\bx{{\mathbf x}}

\newcommand{\be}{\begin{equation}}
\newcommand{\ee}{\end{equation}}
\newcommand{\ba}{\begin{array}}
\newcommand{\ea}{\end{array}}
\newcommand{\bea}{\begin{eqnarray}}
\newcommand{\eea}{\end{eqnarray}}
\newcommand{\beas}{\begin{eqnarray*}}
\newcommand{\eeas}{\end{eqnarray*}}

\numberwithin{equation}{section}

\title{Modulation equations approach for solving vortex and radiation in nonlinear Schr\"{o}dinger equation}

\author{Avy Soffer \thanks{Department of Mathematics, Rutgers University, New Jersey, 08854, USA. School of Mathematics and Statistics, Central China Normal University, Wuhan, 430079, China. ({\tt soffer@math.rutgers.edu})}
 \and Xiaofei Zhao \thanks{Universit\'{e} de Rennes 1, IRMAR, Campus de Beaulieu, 35042 Rennes Cedex, France. ({\tt zhxfnus@gmail.com})}
}

\begin{document}
\maketitle

\begin{abstract}
We apply the modulation theory to study the vortex and radiation solution in the two-dimensional nonlinear Schr\"{o}dinger equation. The full modulation equations which describe the dynamics of the vortex and radiation separately are derived. A general algorithm is proposed to efficiently and accurately find vortices with different values of energy and spin index. The modulation equations are solved by accurate numerical method. Numerical tests and simulations of scattering are given.
 \end{abstract}

\pagestyle{myheadings}
\thispagestyle{plain}
\markboth{A. SOFFER AND X. ZHAO}{Modulation equation for vortex and radiation in NLS}

\section{Introduction}\label{sec:intro}
In our recent work \cite{Zhao1,Zhao2}, we studied the multichannel dynamics of soliton and radiation in the nonlinear Schr\"{o}dinger equation (NLS) by using the modulation equations approach, where the soliton was considered is a ground state. As a kind of soliton with interesting geometric structure admitted by NLS in high space dimensions, vortices have drawn great interest in mathematical analysis and applications \cite{Anderson, Neu,Pego,Cuccagna,Cohen}. In this work, we carry out the investigation by considering the two-dimensional NLS
\begin{equation}\label{nls}
i\partial_tu(t,\bx)=-\Delta u(t,\bx)+V(\bx)u(t,\bx)+\beta(|u(t,\bx)|^2)u(t,\bx),\quad t>0,\ \bx\in\RR^2,
\end{equation}
where $V(\cdot)$ is a radial symmetric real-valued potential function, i.e. $V(\bx)=V(|\bx|)$, and $\beta(\cdot)$ is a real-valued function denoting the nonlinear interaction. 

We study the dynamics of the multichannel solution \cite{Soffer1}: $u=u_{v}+u_{r}$ in the NLS (\ref{nls}), where $u_v$ is a vortex solution and $u_r$ denotes radiation wave.
When there is no radiation, the NLS (\ref{nls}) admits bound state
\begin{equation*}
u_{v}(t,\bx)=\fe^{itw}\phi_w(\bx),\quad \mbox{for some}\  w\in\RR,
\end{equation*}
where $\phi_w(\bx)$ solves the time-independent NLS
\begin{equation}\label{inls}
-\Delta \phi_w+V\phi_w+\beta(|\phi_w|^2)\phi_w=w\phi_w,\quad \bx\in\RR^2.
\end{equation}
The vortex solution is when the function in the bound state takes the form
\begin{equation}\label{form}
\phi_w(\bx)=\fe^{im\theta}\rho_w(r),
\end{equation}
for some $m\in\NN^+$, where $\theta=\arg(x),r=|\bx|$ and $\rho_w(r)$ is a real-valued radial function satisfying the one-dimensional equation resulting from (\ref{inls}):
\begin{equation}\label{rho}
\frac{d^2}{dr^2}\rho_w(r)+\frac{1}{r}\frac{d}{dr}\rho_w(r)-\frac{m^2}{r^2}\rho_w(r)+(V(r)+\beta(\rho_w^2)-w)\rho_w=0,\quad r>0.
\end{equation}
The integer $m\neq0$ is known as the spin index or winding number of the vortex. The vortex (\ref{form}) is also known as the central vortex state in \cite{BaoCai,Markowich} and is a kin to the quantized vortices that occur in superconductivity \cite{tang}. When $m=0$, the solution $u_v$ falls back to the ground state as has been studied in \cite{Zhao1, Zhao2}.  In this sense, the vortex given by (\ref{inls}) and (\ref{rho}) can be interpreted as an excited state among the bound states \cite{Cuccagna}. For given $w$ and $m$, the existence of $\rho_w$ in (\ref{rho}) has been proved in \cite{Warchall} and the solution $\rho_w(r)\to0$ exponentially as $r\to\infty$ \cite{Pego}. Spectrally stable vortices for $m\leq5$ have been found in \cite{Pego}. Orbital instabilities of the vortices have been pointed out and investigated in \cite{Mizumachi,Jones,Pego,Towers,tang} by either mathematical analysis or numerical methods.

The modulation equations which describe the dynamics of the soliton and the radiation part separately were originally introduced in \cite{Soffer1,Soffer2,Weinstein} as an analytical tool for studying scattering and stability of solitons. In recent work \cite{Zhao1,Zhao2}, it has been developed as an alternative approach to solve the multichannel dynamics of the NLS. Compared to the modulation equations approach, classical numerical solvers towards NLS are not able to distinguish the two waves. The popular collective coordinates method \cite{374} would go wrong by completely dropping radiation. On the other hand, the modulation equations approach provides a strategy to design absorbing boundary conditions or filters to filter out the fast dispersing waves \cite{Zhao2}. In \cite{Zhao1,Zhao2,Soffer1}, the modulation equations for the case of a ground state soliton and radiation were studied and the results were emphasized in 1D. To study the case for vortex as an excited soliton in 2D, more efforts need to be devoted. For the computation of the vortex (\ref{form}), solving (\ref{rho}) via numerical approximations is a widely-used approach in the literature \cite{tang,Markowich,Pego,Towers}, since (\ref{rho}) uses the radial symmetric property that reduces the problem to a real-valued one-dimensional equation. However, it also brings some numerical difficulties. A singularity and an artificial boundary condition are introduced in the equation (\ref{rho}) at the origin $r=0$. Using ad hoc boundary approximations as in \cite{Pego} or using a sophisticated polynomial spectral discretization as in \cite{Markowich} would either be less accurate or less efficient. 

In this paper, we first derive the modulation equations for the vortex and radiation. Then we give an iterative algorithm to accurately compute the vortex state (\ref{form}) with prescribed $w$ and $m$ via using (\ref{inls}). After that, we solve the modulation equations for the dynamics of vortex and radiation via some numerical method. Numerical tests are done to validate the proposed approach for NLS and simulations of the scattering of the radiation are given. 

The rest of the paper is organized as follows. In Section \ref{sec:me}, we shall derive the modulation equations. In Section \ref{sec:num}, we shall solve to get the vortex solution and then solve the modulation equations by numerical methods. Numerical results will also be presented in this section. Conclusions are drawn in Section \ref{sec:conc}.

\section{Modulation equations}\label{sec:me}
We define the inner product of two complex-valued functions $f$ and $g$ as
$<f,g>:=\re\{\int_{\RR^2}f(\bx)\overline{g}(\bx)\ud\bx\}$, where $\re(z)$ denote the real part of a complex number.

We make the ansatz \cite{Soffer1,Soffer2,Weinstein} that the solution of a standing vortex and radiation wave in the NLS can be written as
\begin{align}\label{ansatz}
u(t,\bx)=\fe^{-i\alpha(t)}\left(\phi_{w(t)}(\bx)+R(t,\bx)\right),\quad t\geq0,\ x\in\RR^2,
\end{align}
where $\phi_w=\fe^{im\theta}\rho_w(r)$ denotes the vortex satisfying (\ref{inls}) and (\ref{rho}),
\begin{align}\label{alpha}
\alpha(t)=\int_0^tw(s)\ud s-\gamma(t),\quad t\geq0,
\end{align}
for some real-valued function $\gamma$,  and $R$ denotes the radiation wave. Plugging (\ref{ansatz}) into (\ref{nls}), we get
\begin{align}\label{R def}
\pa_tR=&-i\left(-\Delta +V+\beta(|\phi_w|^2)\right)R+i\dot{\alpha}R-i\beta'(|\phi_w|^2)\left(|\phi_w|^2R+(\phi_w)^2\overline{R}\right)-i\dot{\gamma}\phi_w\nonumber\\
&-iN-\dot{w}\pa_w\phi_w,\quad t>0,\ \bx\in\RR^2,
\end{align}
where $N=N(t,\bx)$ denotes the nonlinear part in terms of $R$ coming out from the nonlinearity as
\begin{align*}
N:=\beta(|\phi_{w}+R|^2)(\phi_{w}+R)-\beta(|\phi_w|^2)(\phi_w+R)
-\beta'(|\phi_w|^2)\left(|\phi_w|^2R+(\phi_w)^2\overline{R}\right)
\end{align*}
and $\pa_w\phi_w$ is the solution of the derivative of (\ref{inls}) with respect to $w$ on both sides as
\begin{align*}
\left(-\Delta +V+\beta(|\phi_w|^2)-w\right)\pa_w\phi_w+\beta'(|\phi_w|^2)\left(|\phi_w|^2\pa_w\phi_w+(\phi_w)^2\pa_w\overline{\phi_w}\right)=\phi_w.
\end{align*}
Thanks to the special form (\ref{form}), the above elliptic problem for $\pa_w\phi_w$ can be written as a compact form as
\begin{align}\label{dwphi}
\left(-\Delta +V+\beta(|\phi_w|^2)+2\beta'(|\phi_w|^2)|\phi_w|^2-w\right)\pa_w\phi_w=\phi_w,\quad \bx\in\RR^2.
\end{align}

In order to make (\ref{alpha}) and (\ref{R def}) solvable, we assume the following two orthogonality conditions for all times along with (\ref{ansatz}) as \cite{Cuccagna}:
\begin{align}\label{oc}
<R,\phi_w>=0,\quad <R,i\pa_w\phi_w>=0,\quad t\geq0.
\end{align}
Now taking the inner product on both sides of (\ref{R def}) with $\phi_w$ and then integrating by parts, we get
\begin{align*}
<\pa_tR,\phi_w>=&-<iR,w\phi_w>+\dot{\alpha}<iR,\phi_w>-<g,\phi_w>\nonumber\\
&-\dot{w}<\pa_w\phi_w,\phi_w>,\quad  t>0,
\end{align*}
where
\begin{equation}
g=g(t,\bx):=iN+i\beta'(|\phi_w|^2)\left(|\phi_w|^2R+(\phi_w)^2\overline{R}\right).
\end{equation}
By taking a time derivative of the first orthogonality condition in (\ref{oc}) which gives
$<\pa_tR,\phi_w>=-\dot{w}<R,\pa_w\phi_w>,$
and also noting (\ref{alpha}), the above equation becomes
\begin{align}\label{ode1}
<\pa_w\phi_w,\phi_w-R>\dot{w}+<iR,\phi_w>\dot{\gamma}=-<g,\phi_w>,\quad t>0.
\end{align}
Next, taking the inner product on both sides of (\ref{R def}) with $i\pa_w\phi_w$, we get
\begin{align*}
<\pa_tR,i\pa_w\phi_w>=&-<(-\Delta+V+\beta(|\phi_w|^2))R,\pa_w\phi_w>+\dot{\alpha}<R,\pa_w\phi_w>\\
&-<g,i\pa_w\phi_w>-\dot{\gamma}<\phi_w,\pa_w\phi_w>,\quad t>0.
\end{align*}
By integrating by parts and noting (\ref{dwphi}), we find
$$<(-\Delta+V+\beta(|\phi_w|^2))R,\pa_w\phi_w>=-2<R,\beta'(|\phi_w|^2)|\phi_w|^2\pa_w\phi_w>+w<R,\pa_w\phi_w>.$$
Together by using the second orthogonality condition in (\ref{oc}), we can further get
\begin{align}\label{ode2}
-<R,i\pa_w^2\phi_w>\dot{w}+<\phi_w+R,\pa_w\phi_w>\dot{\gamma}=<f,\pa_w\phi_w>,\quad t>0,
\end{align}
where
\begin{equation}
f:=\beta'(|\phi_w|^2)\left(|\phi_w|^2R-(\phi_w)^2\overline{R}\right)-N,
\end{equation}
and $\pa_w^2\phi_w$ is the solution of the elliptic problem
\begin{align}
&\quad\left[-\Delta +V+\beta(|\phi_w|^2)+2\beta'(|\phi_w|^2)|\phi_w|^2-w\right]\pa_w^2\phi_w\nonumber\\
&=2\pa_w\phi_w-
6\beta'(|\phi_w|^2)|\pa_w\phi_w|^2\phi_w-4\beta''(|\phi_w|^2)|\phi_w\pa_w\phi_w|^2\phi_w,\quad \bx\in\RR^2.\label{dw2phi}
\end{align}

Now defining
\begin{equation}
 \displaystyle A(t):=\left(\begin{matrix}
<\pa_w\phi_w,\phi_w-R> & & <iR,\phi_w>\\
-<R,i\partial_w^2\phi_w>&&<\pa_w\phi_w,\phi_w+R>
\end{matrix}\right),\quad G(t):=\binom{-<g,\phi_w>}{<f,\pa_w\phi_w>},
\end{equation}
and combining (\ref{ode1}), (\ref{ode2}) and (\ref{R def}), we get the full modulation equations for describing the dynamics of a standing vortex and radiation as
\begin{subequations}\label{me}
\begin{align}
&A(t)\binom{\dot{w}}{\dot{\gamma}}=G(t),\quad t>0,\\
&\pa_tR=-i\left(-\Delta +V+\beta(|\phi_w|^2)\right)R+i(w-\dot{\gamma})R-g-i\dot{\gamma}\phi_w\nonumber\\
&\qquad\ \ -\dot{w}\pa_w\phi_w,\quad t>0,\ \bx\in\RR^2.\label{me R}
\end{align}
\end{subequations}
$w$ and $\gamma$ are sometimes referred as collective coordinates in the literature. 
We remark that the equation (\ref{me R}) could be written as a two-component real-valued system form by separating the real part and imaginary part in the equation, which is widely used in the literature \cite{Pego,Cuccagna,Zhao2} for analysis. However, the real part or imaginary part would involve the angle $\theta$  which is undefined at origin and would be a problem for the computer. Here we use the complex-valued single equation form (\ref{me R}) for the convenience of forthcoming numerical approximations. 

Suppose initially the multichannel solution to the NLS (\ref{nls}) is given by
$$u(0,\bx)=\fe^{i\gamma_0}\left(\phi_{w_0}(\bx)+R_0(\bx)\right),\quad x\in\RR^2,$$
then the modulation equations (\ref{me}) are assigned with initial data 
\begin{equation}\label{me ini}
w(0)=w_0,\quad \gamma(0)=\gamma_0,\quad R(0,\bx)=R_0(\bx).
\end{equation}
If we assume initially the vortex and radiation satisfying the orthogonality condition (\ref{oc}), i.e.
$$<R_0,\pa_{w_0}\phi_w>=<R_0,i\pa_w\phi_{w_0}>=0,$$
then the initial value problem of (\ref{me}) and (\ref{me ini}) is well-posed for orbitally stable vortices and the solution given by (\ref{ansatz}) solves the NLS (\ref{nls}).

\section{Numerical method and result}\label{sec:num}
In this section, we shall apply some accurate numerical methods to solve the time-independent NLS for vortex and then solve the modulation equations for multischannel dynamics.

\subsection{Solving the elliptic problem for vortex}
Obtaining the vortices (\ref{form}) accurately and efficiently from the time-independent NLS (\ref{inls}) or (\ref{rho}) with prescribed $w$ and $m$ is an interesting numerical approximation issue, which is also a fundamental step here for solving the modulation equations (\ref{me}) during the dynamics.

For ground state solution (positive $\phi_w$ or saying $m=0$), the Petviashvili method works well \cite{Lakoba,Petviashvili,Pelinovsky}, while vortices bound state need some other treatments. Based on the background of the vortices, different approximation methods have been proposed in the literature. For instance, in studies of rotating Bose-Einstein condensates (BEC), vortices patterns will form in the ground states (energy minimizer) \cite{BaoCai} according to the rotating velocity and the vortices formed are all with the stable index $m=1$. The normalized gradient flow method \cite{BaoCai} or the Krylov subspace method \cite{Xavier} is the powerful numerical method for computing those vortices patterns. For obtaining the vortex (\ref{form}) which is also known as the center vortex state \cite{Markowich}, there are basically two kinds of methods available in the literature as far as we know. One is solving (\ref{rho}) by the shooting method with ad hoc approximations at the boundaries, i.e. $r=0$ and $r=L$ (with $L$ some sufficiently large value). For example, the asymptotic behavior of Bessel functions are used in \cite{Pego}. This kind of method is restricted by its accuracy due to the ad hoc boundary condition and the singularity in the equation (\ref{rho}) at $r=0$. On the other hand, for solving (\ref{rho}) as well, a generalized-Laguerre pseudospectral method has been developed in \cite{BaoCai,Markowich}. This method is very accurate. However, the polynomial pseudospectral method has no fast transformations in practice, and it is not easy to program.

Here we solve the vortex (\ref{form}) based on the equation (\ref{inls}) instead of making use of (\ref{rho}). Comparing to (\ref{rho}), (\ref{inls}) has no singularities in the equation and no artificial boundary condition needed to be specified at the origin. In order to create some non-zero angular momentum in the equation (\ref{inls}) for producing a vortex, we mimic the rotating BEC and introduce the operator
$$L_z=-i(x\pa_y-y\pa_x)=-i\pa_\theta,$$
which allows us to transform (\ref{inls}) into
\begin{equation}
-\Delta \phi_w+V\phi_w+\beta(|\phi_w|^2)\phi_w=\frac{w}{m}L_z\phi_w.
\end{equation}
Then we apply the iterative strategy proposed in \cite{Zhao1,Zhao2}. Thus, for given $w\in\RR,m\in\NN^+$, the detailed algorithm reads as:

\emph{Step 1.} Choose an initial guess $\phi_w^0$.

\emph{Step 2.} For $n\geq0$, find the ground state of the linearized Hamiltonian functional
\begin{equation*}
H^n(\phi)=\int_{\RR^2}\left[|\nabla\phi|^2+V|\phi|^2+\beta\left(|\phi_w^n|^2\right)|\phi|^2-\Omega \overline{\phi}L_z\phi\right]\ud \mathbf{x},
\end{equation*}
in the unit sphere of $L^2(\RR^2)$, where
$\Omega=\frac{w}{m}.$
 Denote the solution as
\begin{equation}\label{algor: eigen1}
\widetilde{\phi}_w^{n+1}:= \arg \min\{H^n(\phi):\phi\in L^2(\RR^2),\ \|\phi\|_{L^2}=1\}.
\end{equation}

\emph{Step 3.} Scale the ground state $\widetilde{\phi}_w^{n+1}$ according to the energy $w$. That is to find the scaling constant $c^n\in\RR$ such that
\begin{equation}\label{algor: eigen2}
\phi_w^{n+1}:=c^n\widetilde{\phi}_w^{n+1},
\end{equation}
satisfying
\begin{align*}
\int_{\RR^2}\left[|\nabla\phi_w^{n+1}|^2+V|\phi_w^{n+1}|^2+\beta\left(|\phi_w^{n+1}|^2\right)|\phi_w^{n+1}|^2\right]\ud \mathbf{x}=w\left\|\phi_w^{n+1}\right\|_{L^2}^2,
\end{align*}
which is obtained by taking the inner product of (\ref{inls}) on both sides with $\phi_w$ in $L^2(\RR^2)$. Then we can solve the equation
$$\int_{\RR^2}\beta\left(\left|c^n\widetilde{\phi}_w^{n+1}\right|^2\right)|\widetilde{\phi}_w^{n+1}|^2\ud \bx=w-\int_{\RR^2} \left(\left|\nabla\widetilde{\phi}_\omega^{n+1}\right|^2+V\left|\widetilde{\phi}_w^{n+1}\right|^2\right)\ud \bx,$$
for the value of $c^n$. In particular, when it comes to the power nonlinearity (or polynomial type like the focusing-defocusing nonlinearity \cite{Pego}) case, i.e. $\beta(\rho)=\lambda\rho^{p},\rho\in\RR,\lambda\in\RR,p>0$, we have explicit  formula
\begin{equation}
c^n=\left|\frac{w-\int_{\RR^2} \left(\left|\nabla\widetilde{\phi}_\omega^{n+1}\right|^2+V|\widetilde{\phi}_w^{n+1}|^2\right)\ud \bx}
{\lambda\int_{\RR^2}\left|\widetilde{\phi}_w^{n+1}\right|^{2p+2}\ud\mathbf{x}}\right|^{\frac{1}{2p}}.
\end{equation}
Then iterate until $\{\phi^n_w\}_{n\geq0}$ converges. The Cauchy criterion can be used as the stopping condition, i.e.
\begin{equation}\label{Cauchy1}
\|\phi^{n+1}_\omega-\phi^n_\omega\|_{L^\infty}\leq \eps,
\end{equation}
with some chosen threshold $\eps>0,$ or
we can measure the residue of the time-independent NLS (\ref{inls}) by defining the error
$$e_{res}(\bx)=-\Delta\phi_w+V\phi_w+\beta(|\phi_w|^2)\phi_w-w\phi_w,\quad \bx\in\RR^2,$$
and stop at the desired threshold, i.e. $\|e_{res}\|_{L^\infty}\leq\eps.$

Here we make some important remarks on the choice of initial guess $\phi_w^0$ in the algorithm.
For $m=1$, the initial guess could be chosen as $\phi_w^0=(x+iy)\fe^{-x^2-y^2}$ as suggested for rotating BEC in \cite{BaoCai} for faster convergence. For integer $m>1$, we need to take an initial guess with the same phase part as the vortex (\ref{form}) in order to create the extra potential energy contributed from the phase. A simple choice could be 
\begin{equation}
\phi_w^0=\frac{(\phi_w)^m}{\|\phi_w\|_{L^\infty}^m},
\end{equation}
 where $\phi_w$ is a vortex solution corresponding to $m=1$.

For implementation issue, the whole space $\RR^2$ in above will be truncated to a bounded interval $I=[a,b]\times[c,d]$ which is large enough for the vortex, with periodic or zero boundaries. The space will be discretized by Fourier pseudospectral method \cite{Shen}. The ground state solver in Step 2 could either be the normalized gradient flow or the Krylov method. Once the vortex $\phi_w$ is obtained, we can discretize (\ref{dwphi}) and (\ref{dw2phi}) into linear systems with the help of the Fourier pseudospectral method for spatial approximations, and then solve the linear systems, say by GMRES for $\pa_w\phi_w$ and $\pa_w^2\phi_w$, respectively. We shall take the cubic nonlinearity, i.e. $\beta(\rho)=\lambda\rho,$ for some $\lambda\in\RR$ in the following numerical examples for simplicity.

\begin{figure}[t!]
{$$
\begin{array}{ccc}
\psfig{figure=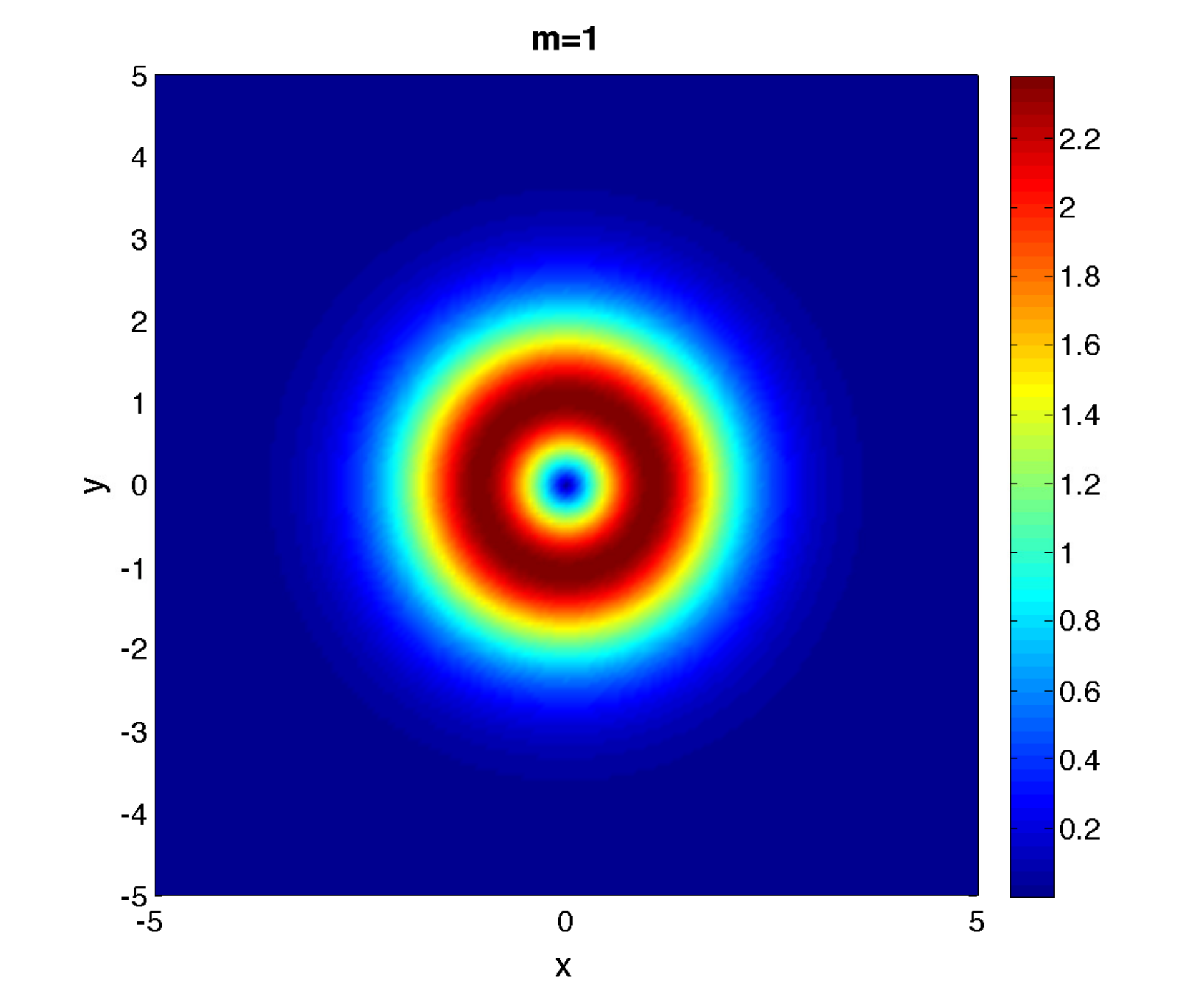,height=4cm,width=4cm}&\psfig{figure=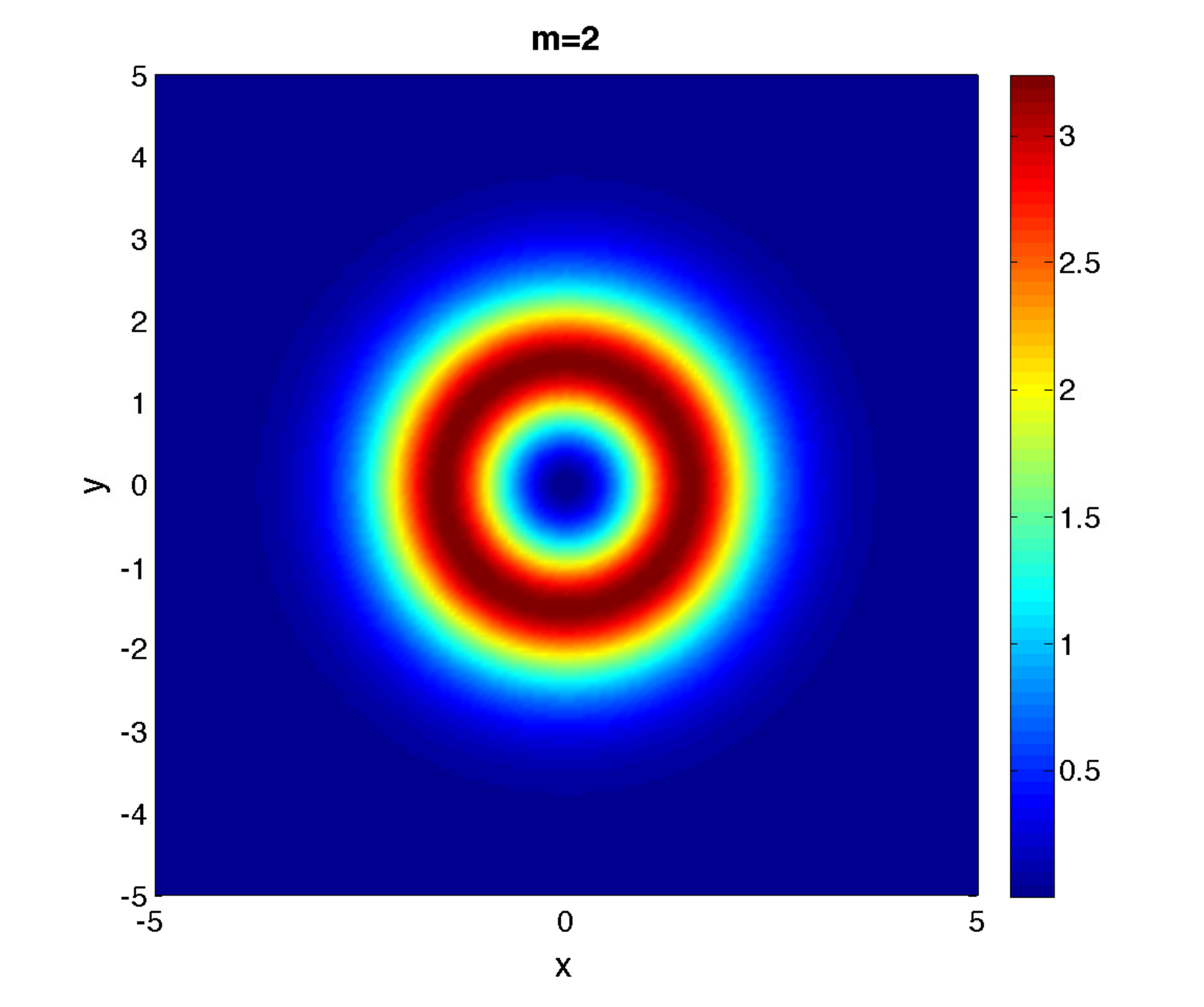,height=4cm,width=4cm}&\psfig{figure=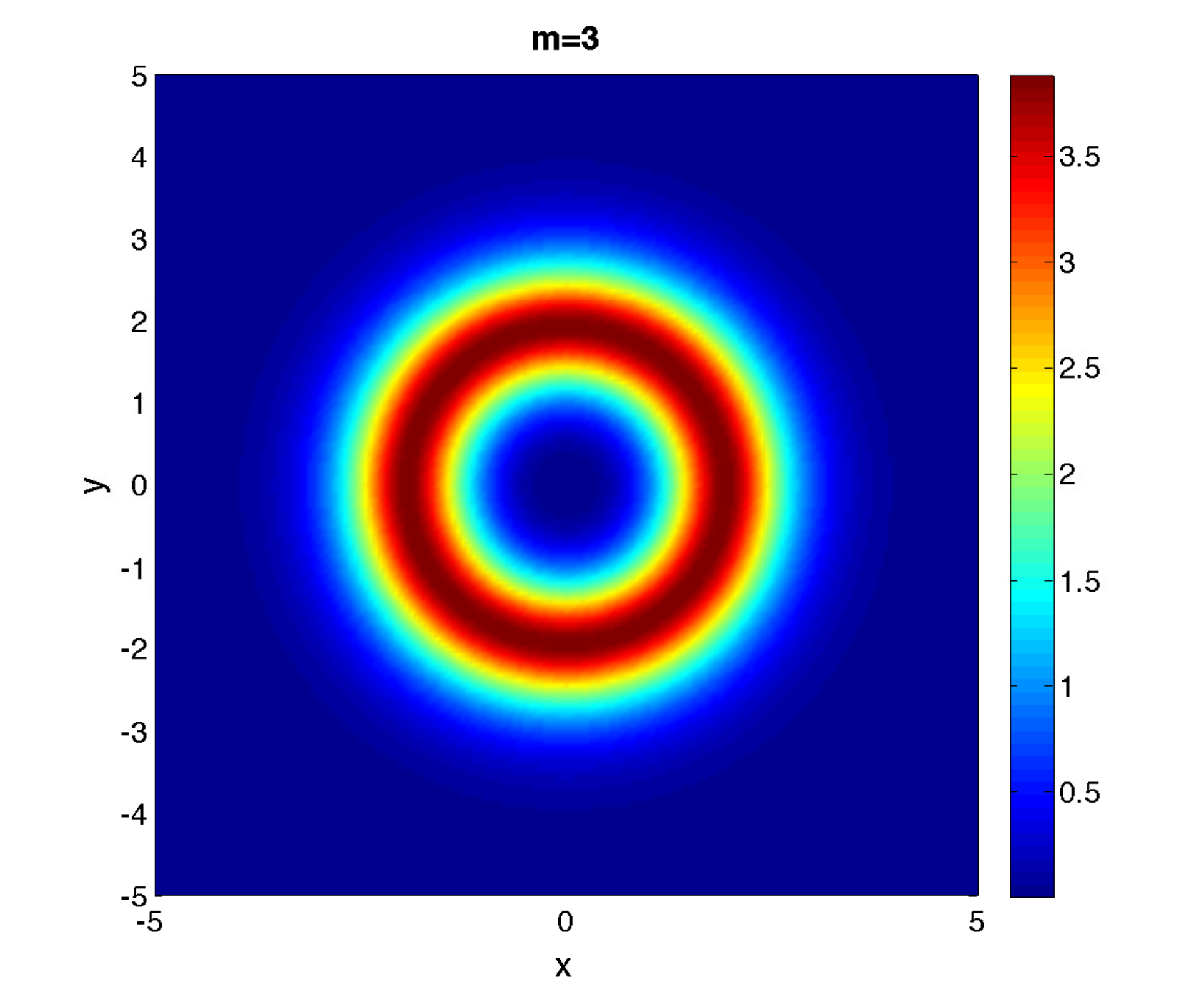,height=4cm,width=4cm}\\
\psfig{figure=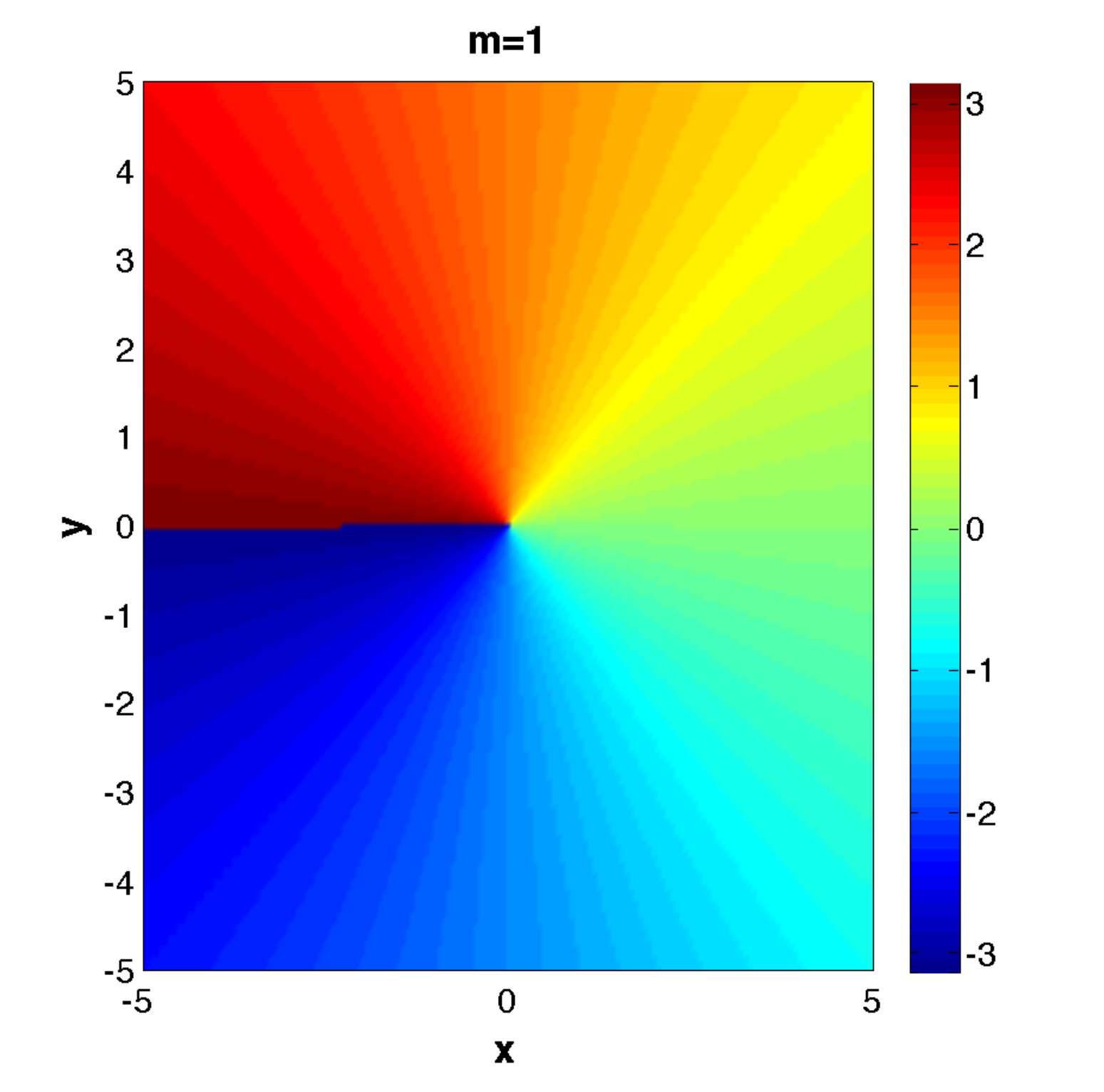,height=4cm,width=4cm}&\psfig{figure=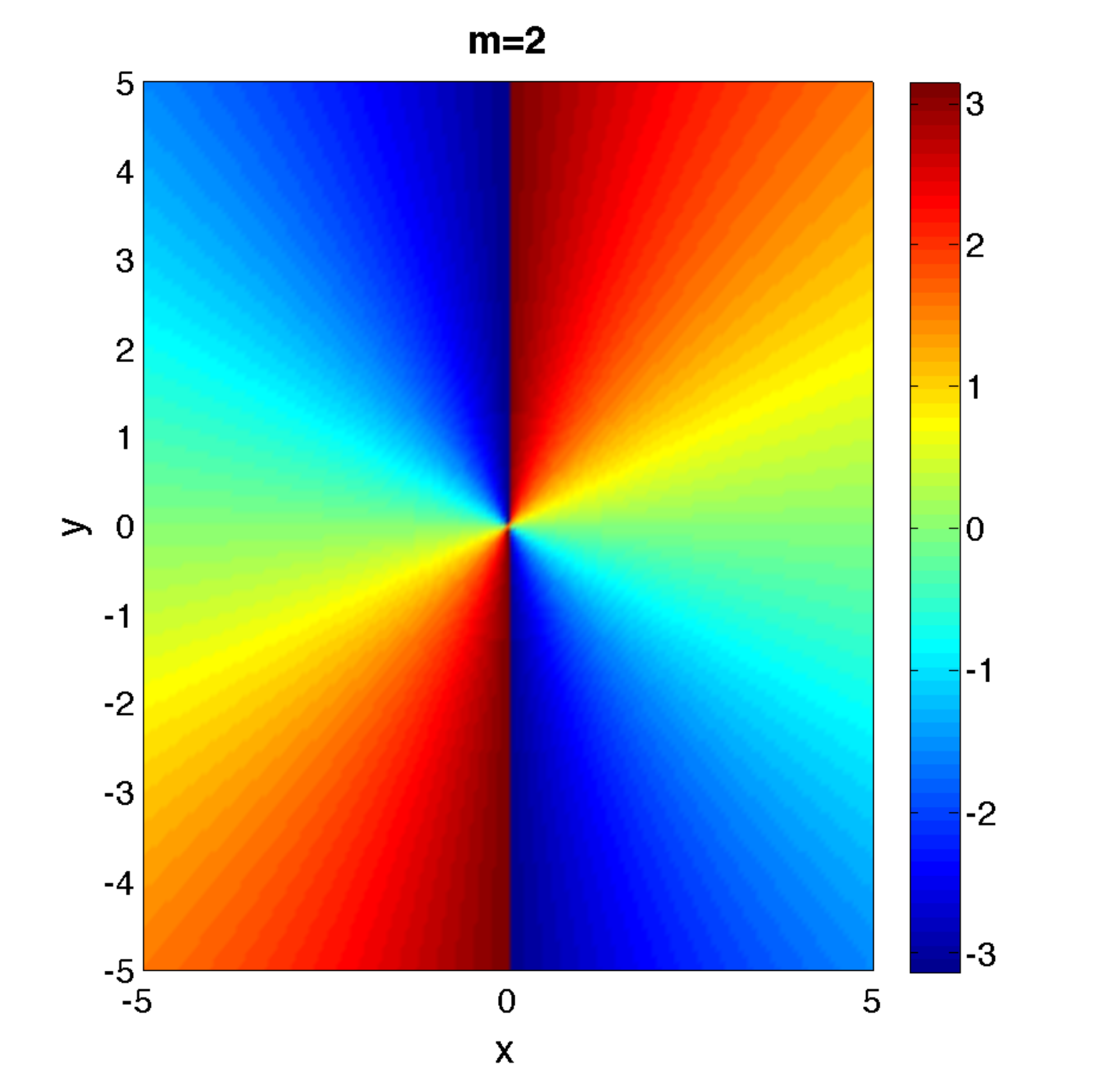,height=4cm,width=4cm}&\psfig{figure=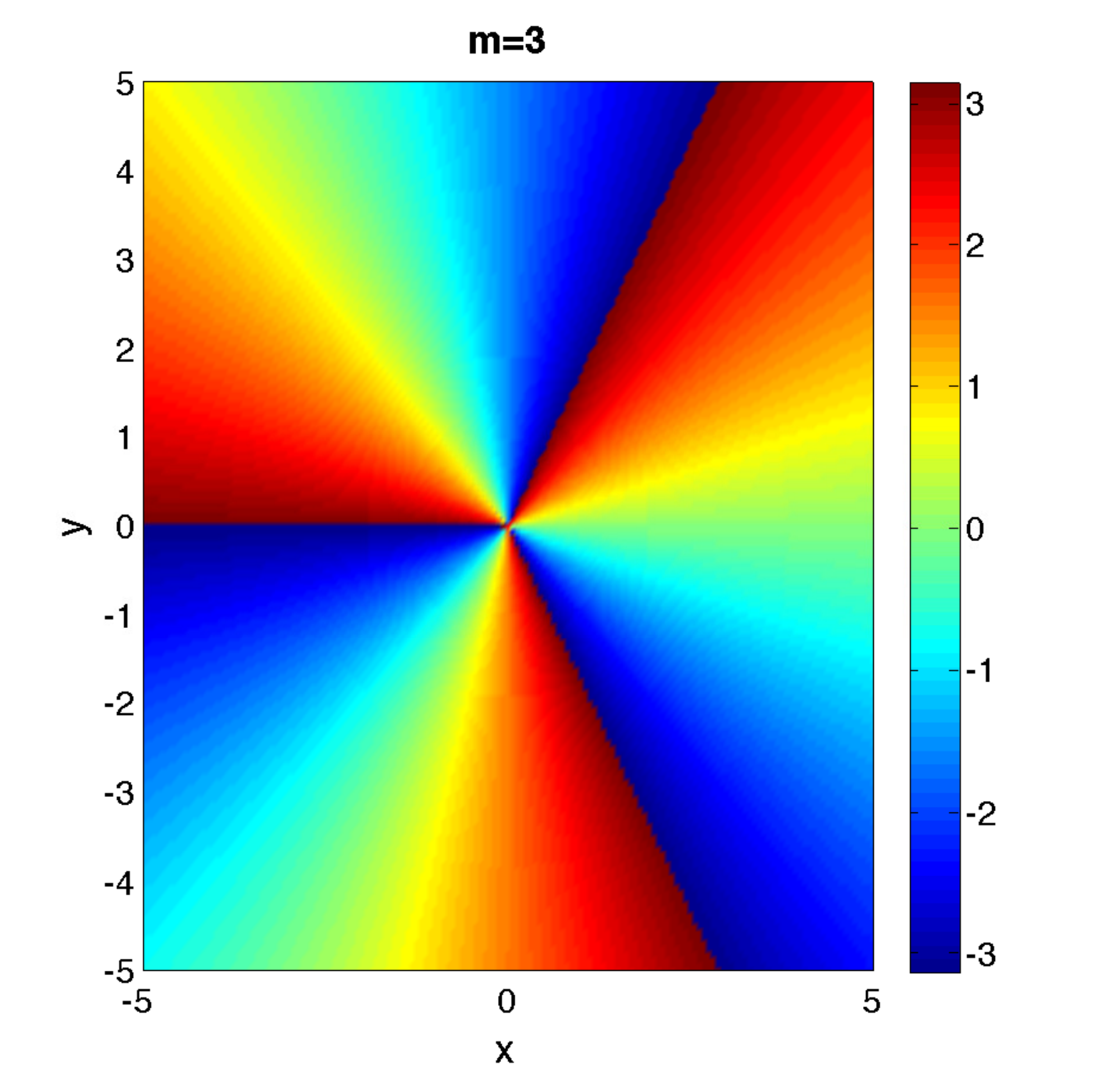,height=4cm,width=4cm}
\end{array}
$$
}
\caption{Contour profiles of vortices in example \ref{vor:eg1} with index $m=1,2,3$: $|\phi_w(x,y)|$ (first row) and $\arg(\phi_w(x,y))$ (second row). }\label{fig:vortex}
\end{figure}

\begin{figure}[t!]
{$$
\begin{array}{cc}
\psfig{figure=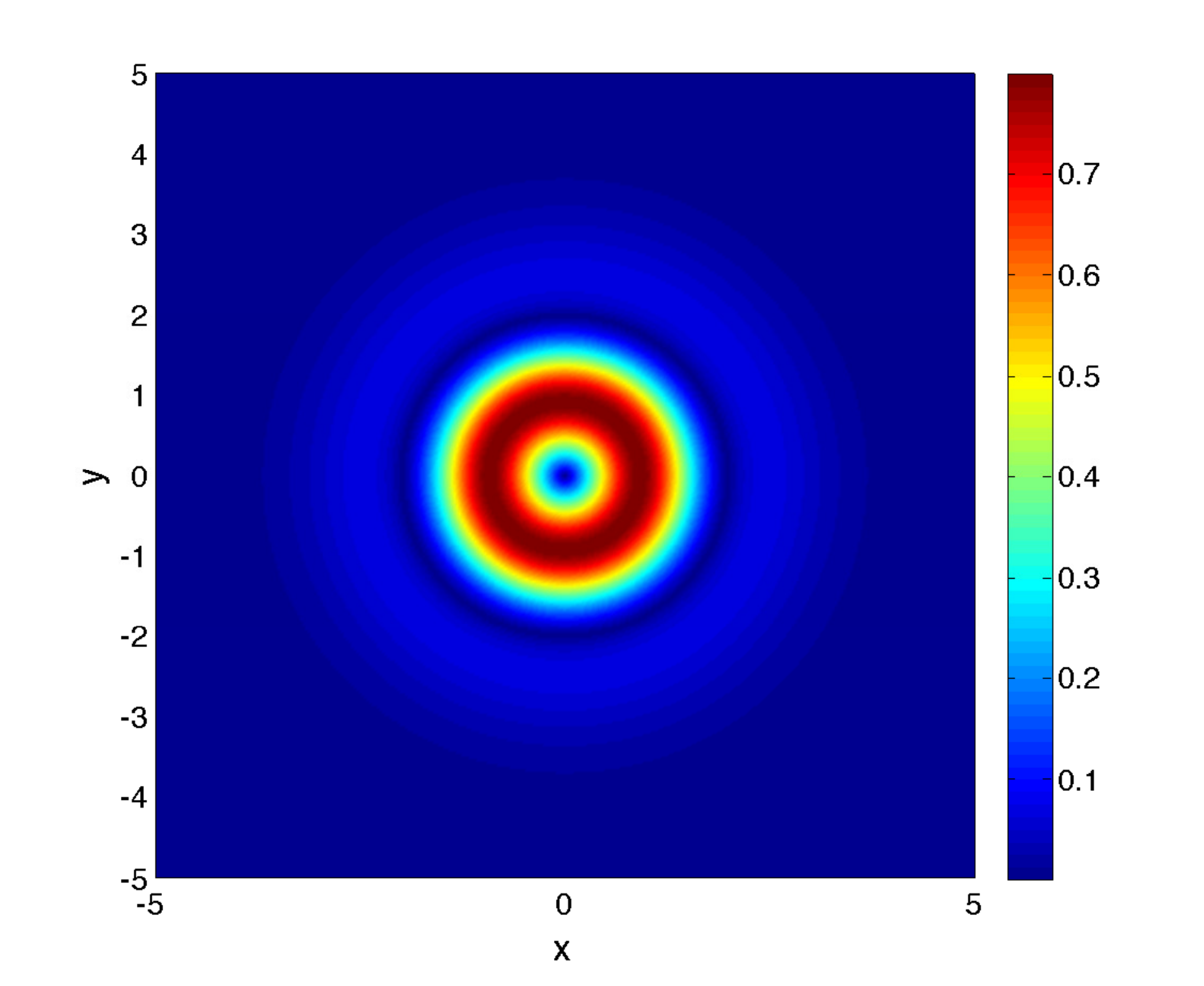,height=4.5cm,width=5.5cm}&\psfig{figure=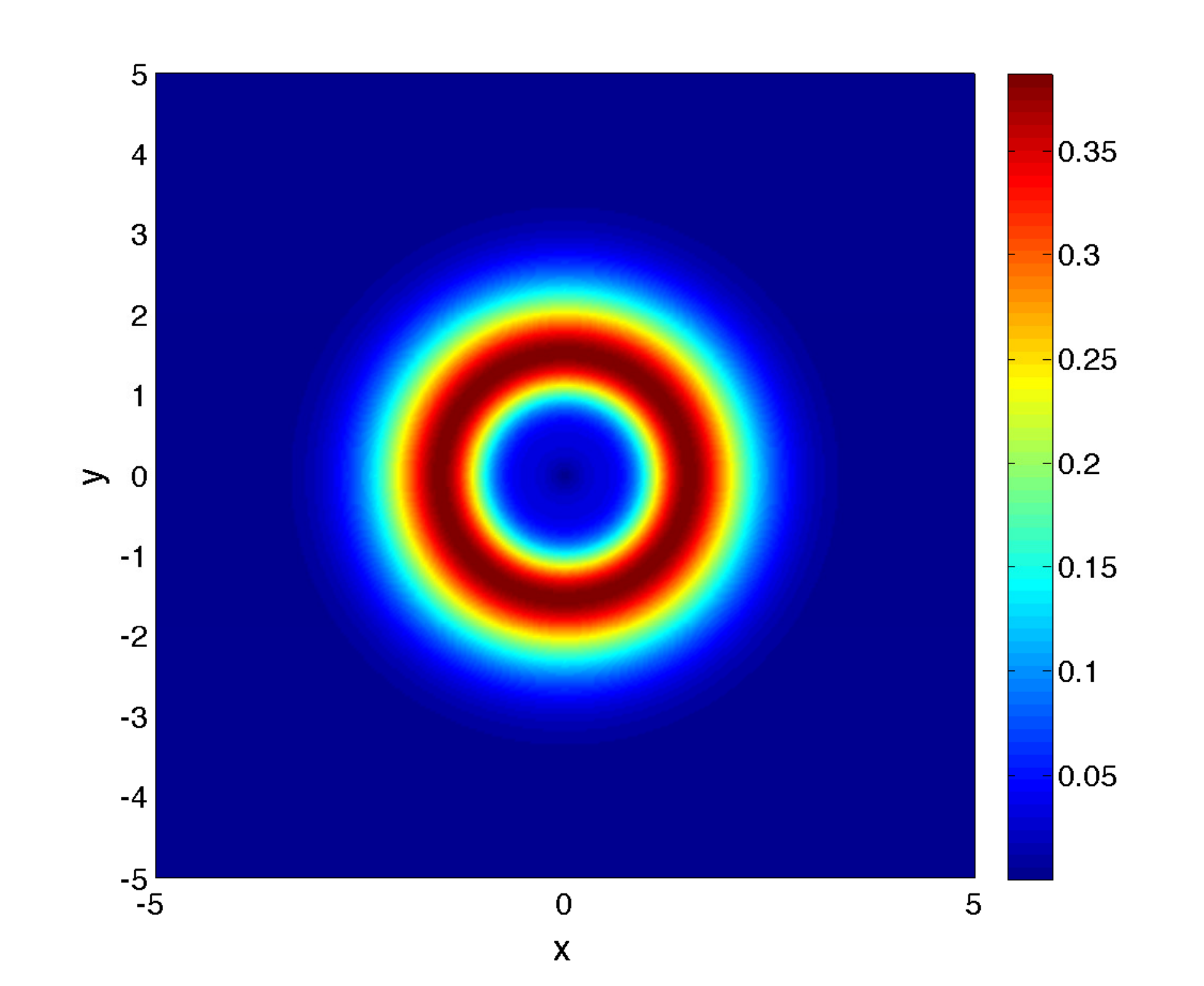,height=4.5cm,width=5.5cm}
\end{array}
$$
}
\caption{Contour plot of $|\pa_w\phi_w(x,y)$ (left) and $|\pa_w^2\phi_w(x,y)|$ (right) in example (\ref{vor:eg1}) with $m=1$. }\label{fig:dwphi}
\end{figure}

In our first numerical example, we consider the vortex that appears in Gross-Pitaevskii equation for rotating BEC \cite{BaoCai} by choosing
\begin{equation}\label{vor:eg1}
V(\bx)=\frac{|\bx|^2}{2},\quad \lambda=-0.5,\quad w=1.1.
\end{equation}
The problem is solved on a bounded interval $I=[-8,8]\times[-8,8]$. $\|e_{res}\|_{L^\infty}\leq\eps$ with several values of $\eps$ are used as the stopping condition.
The convergence results of the iterative algorithm including the residue $e_{res}$, the total number of iterations $n_{tol}$ and the Cauchy error $e_{c}:=\phi_w^{n_{tol}}-\phi_w^{n_{tol}-1}$ for obtaining the vortex in (\ref{vor:eg1}) with spin index $m=1$ are tabulated in Table \ref{tab:vortex}. The profiles of the solutions, i.e. $|\phi_w|$ and $\arg(\phi_w)$ with the index $m=1,2,3$ are plotted in Figure \ref{fig:vortex}. The corresponding profiles of $\pa_w\phi_w$ and $\pa_w^2\phi_w$ for $m=1$ are shown in Figure \ref{fig:dwphi}.

\begin{table}[t!]
  \caption{Convergence of the iterative algorithm for vortex in (\ref{vor:eg1}).}\label{tab:vortex}
  \vspace*{-10pt}
\begin{center}
\def\temptablewidth{1\textwidth}
{\rule{\temptablewidth}{0.75pt}}
\begin{tabular*}{\temptablewidth}{@{\extracolsep{\fill}}llllll}
                                 & $\eps=0.1$       &  $\eps=0.05$       & $\eps=0.01$           & $\eps=0.005$ \\[0.25em]
\hline
$\left\|e_{res}\right\|_{L^\infty}$  &5.62E-2	   &3.43E-2	        &9.90E-3	        &5.00E-3      \\
$\left\|e_{c}\right\|_{L^\infty}$                            &9.60E-3       &4.80E-3         &3.76E-5            &1.85E-5\\
$n_{tol}$                      &6              &7               &37                 &89
\end{tabular*}
{\rule{\temptablewidth}{0.75pt}}
\end{center}
\end{table}


\begin{figure}[h!]
{$$
\begin{array}{ccc}
\psfig{figure=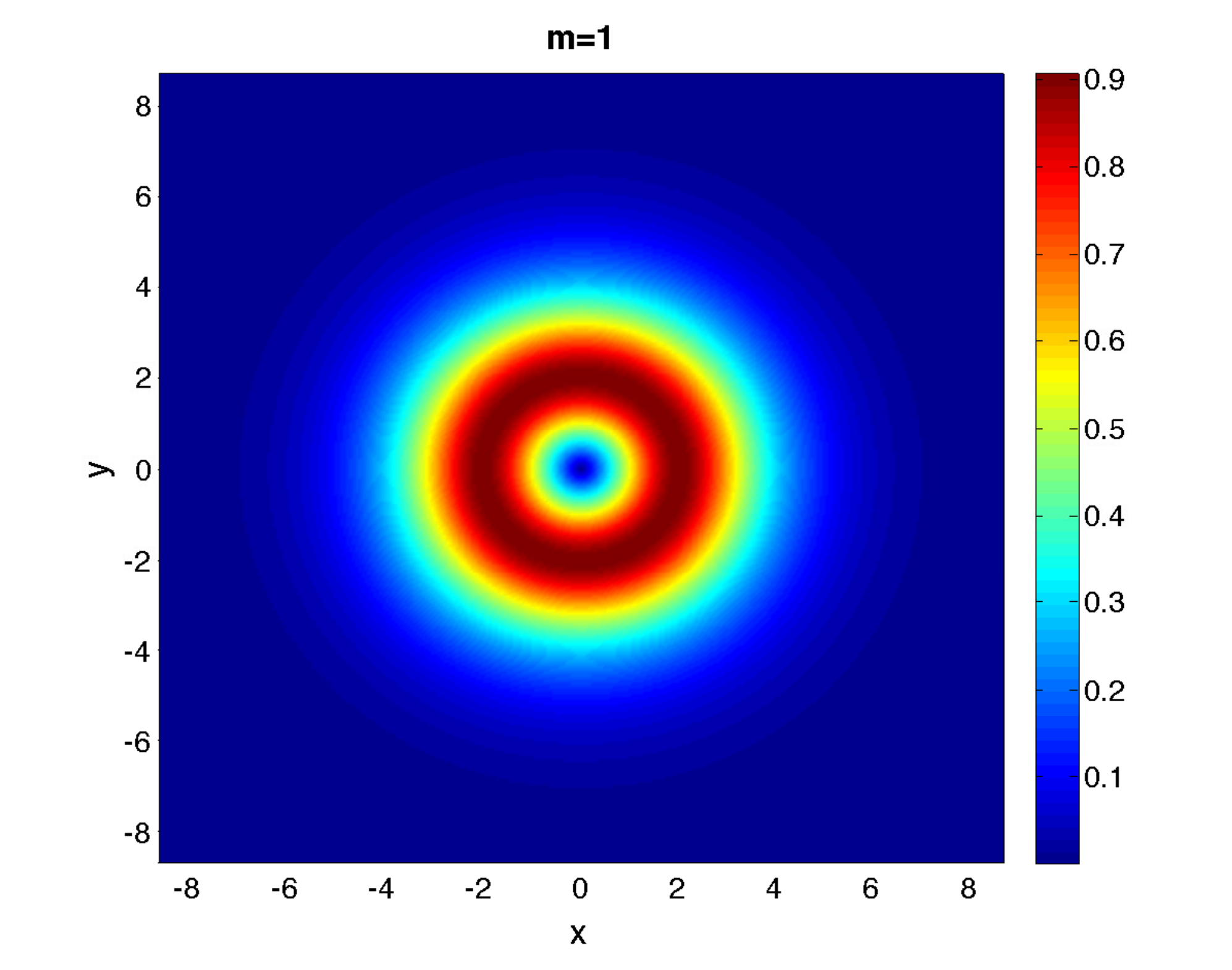,height=4cm,width=4cm}&\psfig{figure=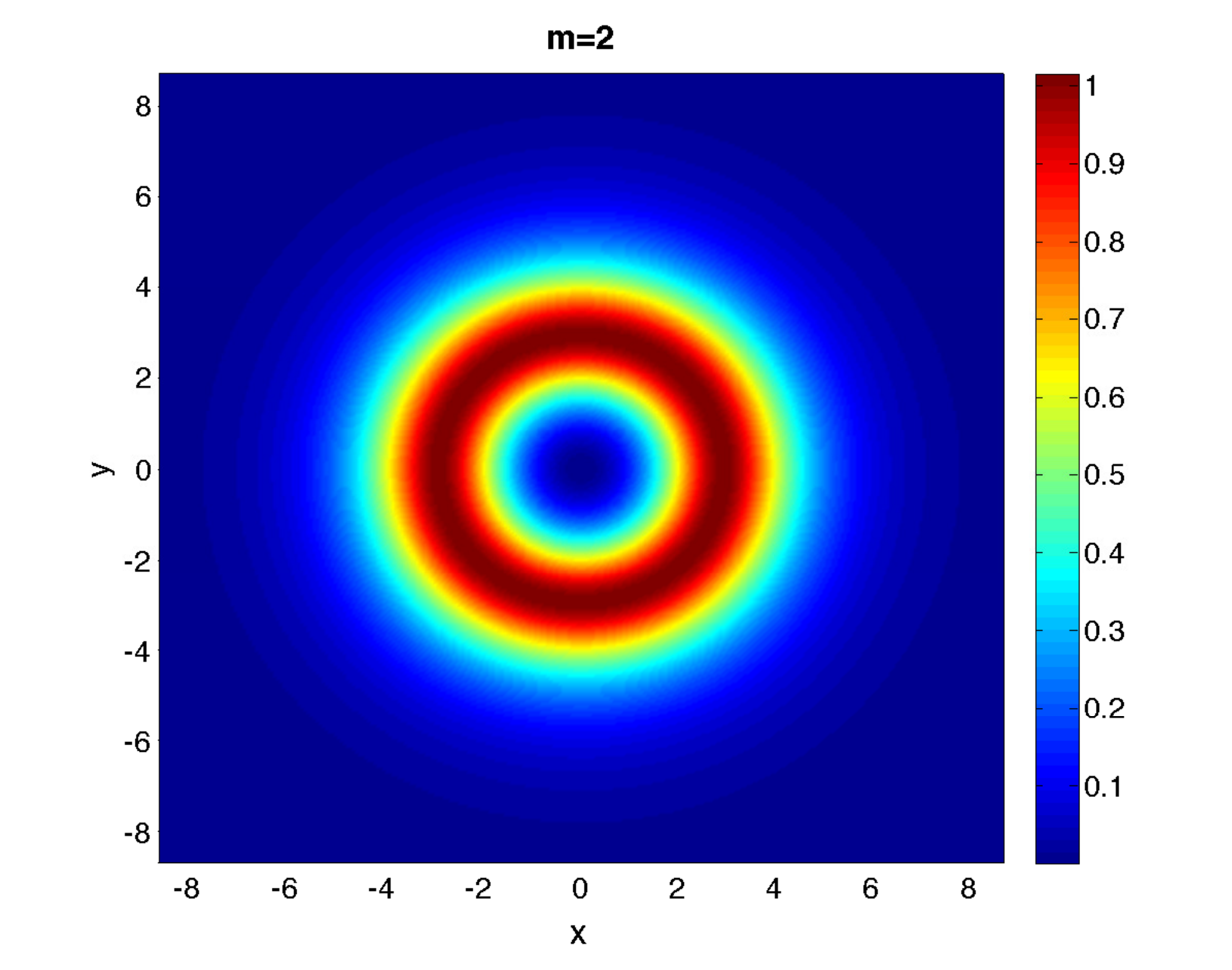,height=4cm,width=4cm}&\psfig{figure=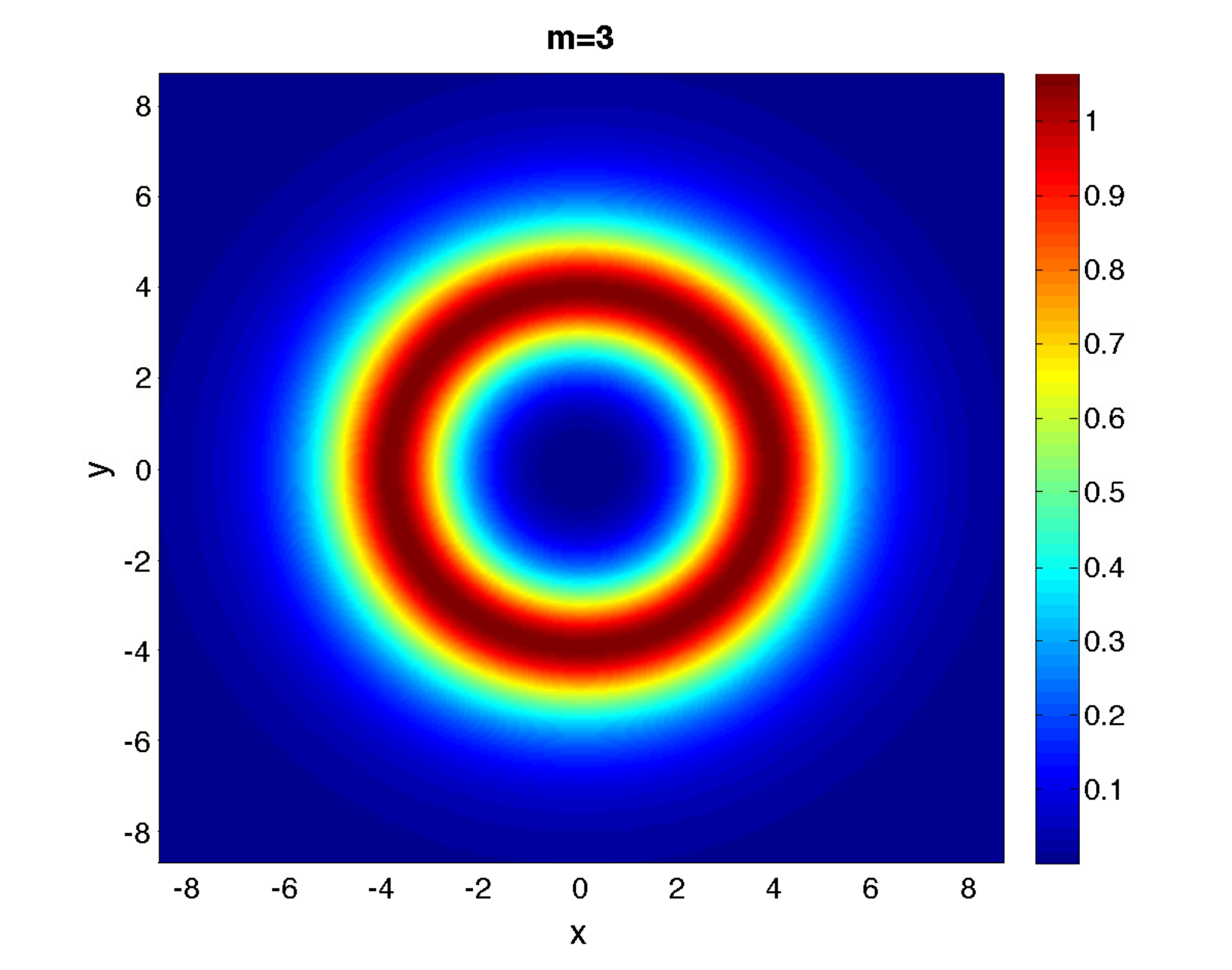,height=4cm,width=4cm}\\
\psfig{figure=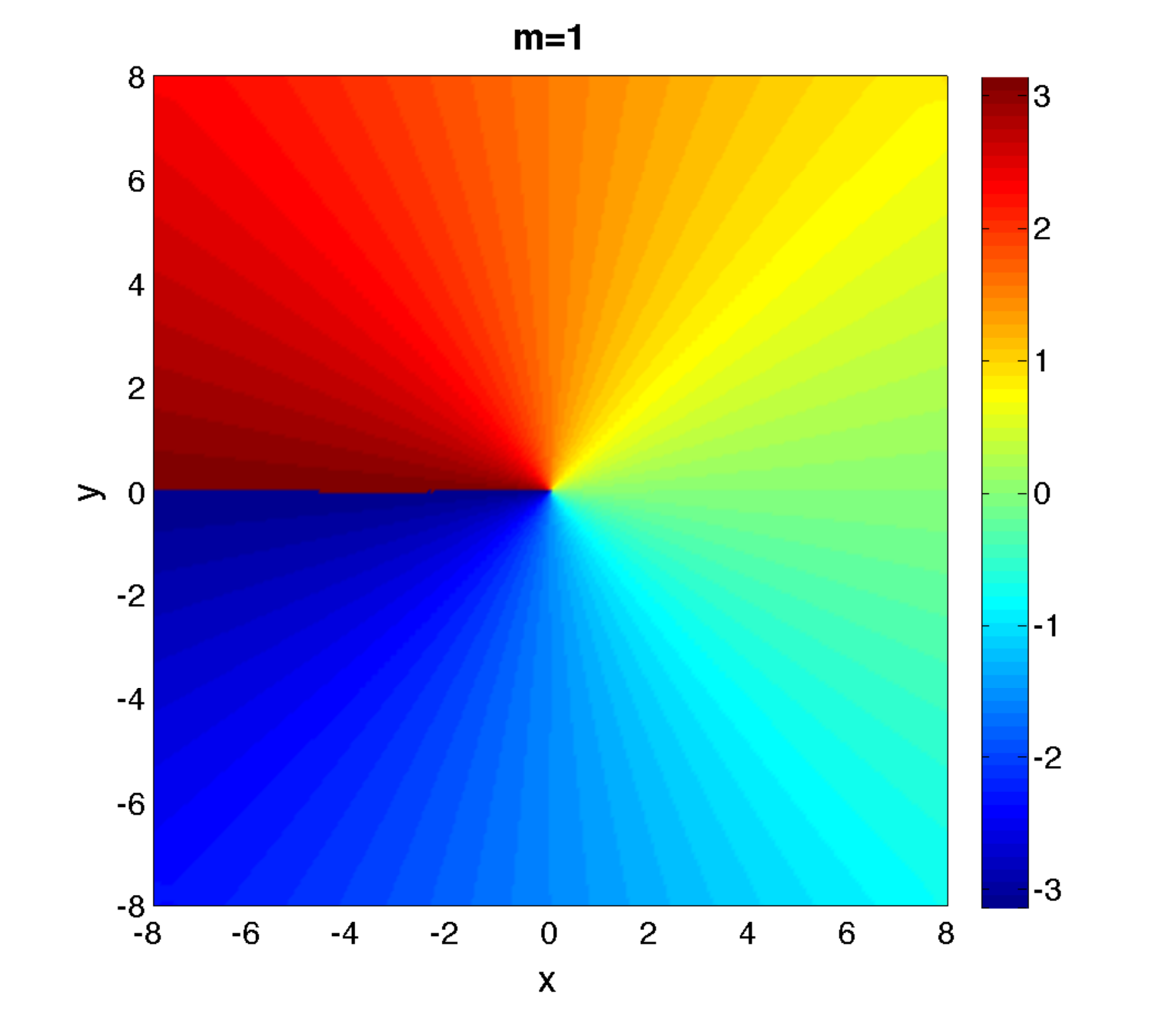,height=4cm,width=4cm}&\psfig{figure=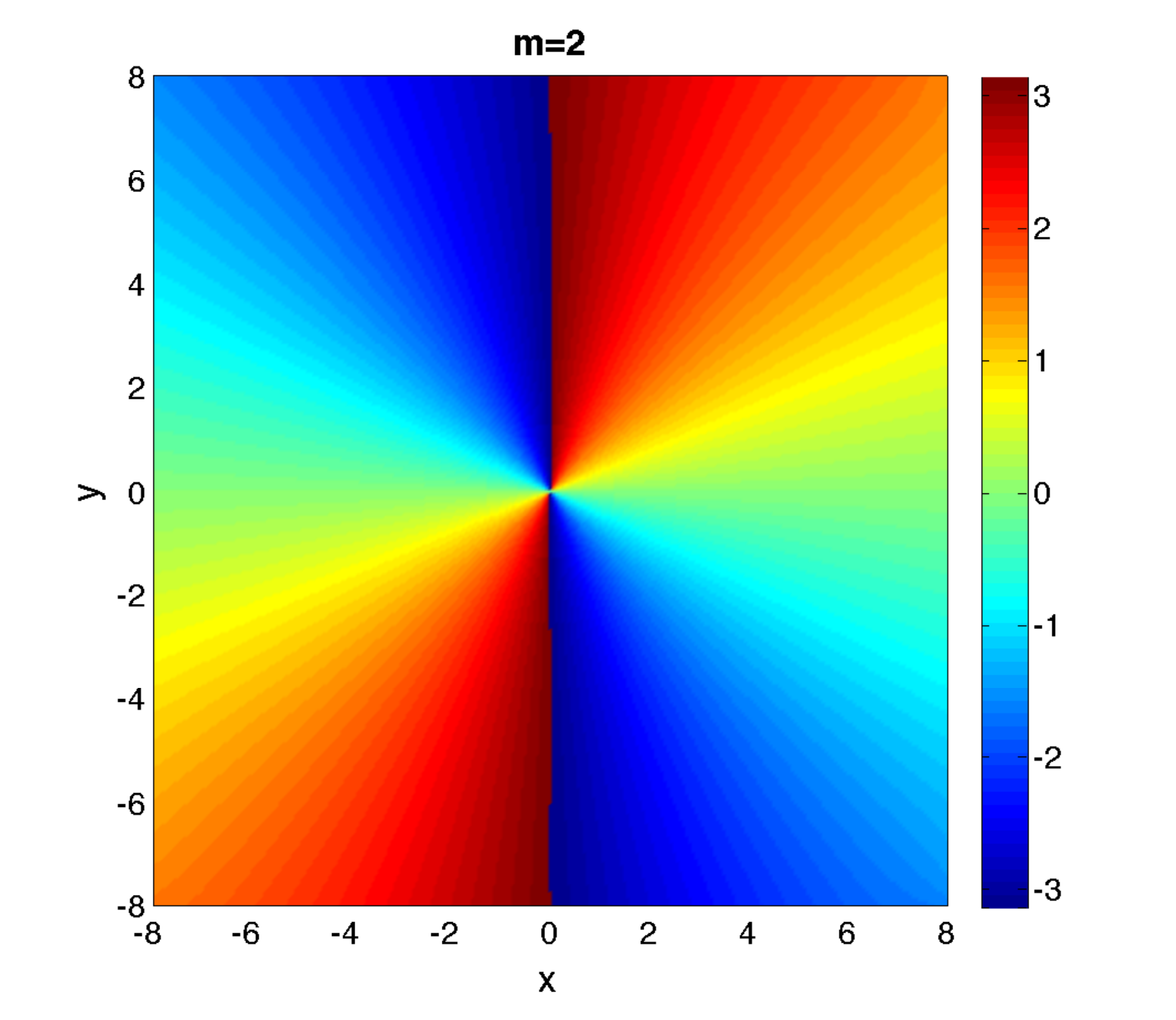,height=4cm,width=4cm}&\psfig{figure=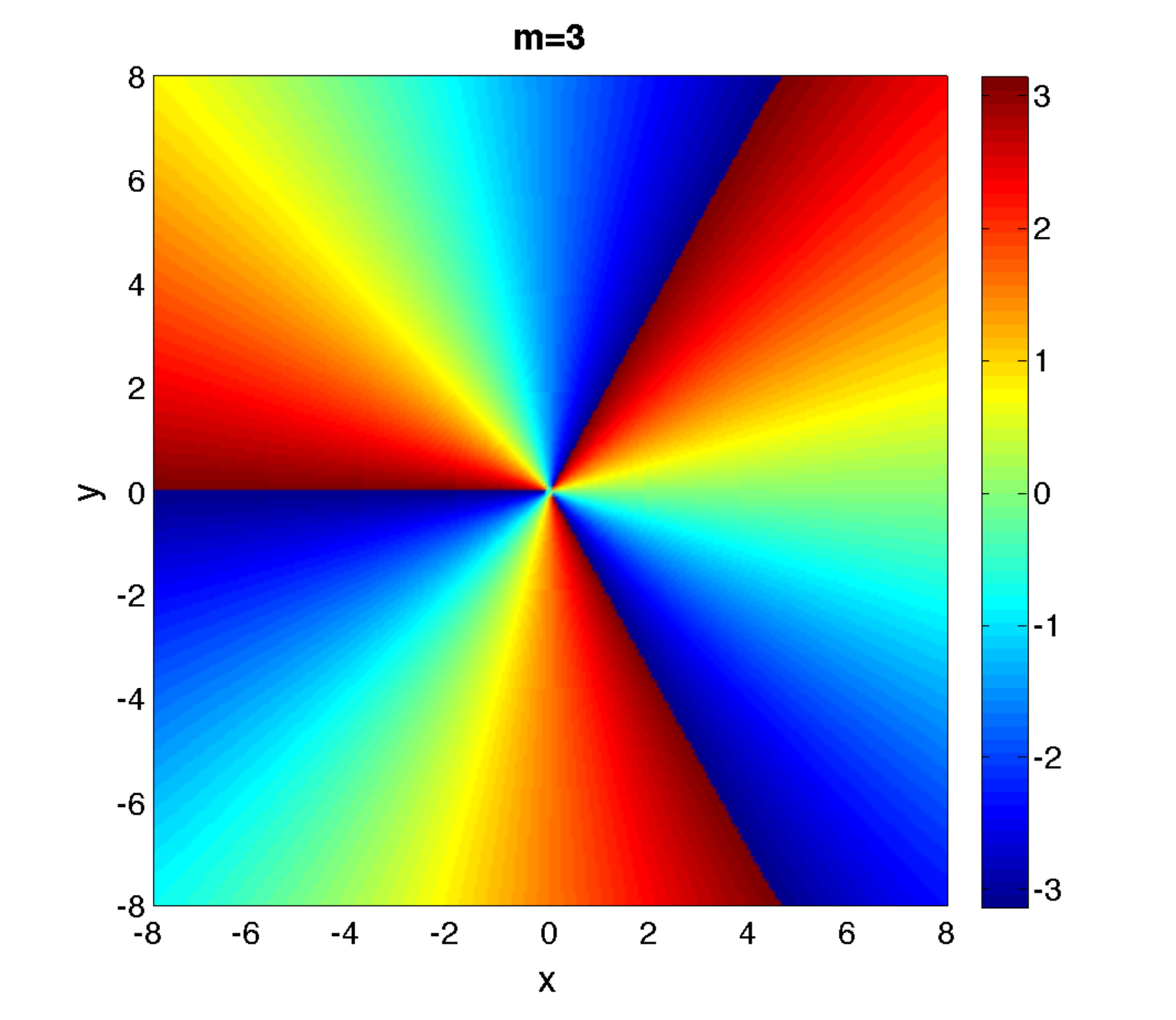,height=4cm,width=4cm}
\end{array}
$$
}
\caption{Contour profiles of vortices in example \ref{vor:eg2} with index $m=1,2,3$: $|\phi_w(x,y)|$ (first row) and $\arg(\phi_w(x,y))$ (second row). }\label{fig:vortex2}
\end{figure}

In our second example, we consider the vortex that occurs in nonlinear optics \cite{Pego} by choosing
\begin{equation}\label{vor:eg2}
V(\bx)=0,\quad \lambda=-2,\quad w=-0.5.
\end{equation}
The problem is solved on interval $I=[-12,12]\times[-12,12]$.
The corresponding convergence results in case $m=1$ are given in Table \ref{tab:vortex2} and the profiles of the vortices for $m=1,2,3$ are shown in Figure \ref{fig:vortex2}.

\begin{table}[t!]
  \caption{Convergence of the iterative algorithm for vortex in (\ref{vor:eg2}).}\label{tab:vortex2}
  \vspace*{-10pt}
\begin{center}
\def\temptablewidth{1\textwidth}
{\rule{\temptablewidth}{0.75pt}}
\begin{tabular*}{\temptablewidth}{@{\extracolsep{\fill}}llllll}
                                 & $\eps=0.1$       &  $\eps/2$       & $\eps/4$           & $\eps/8$ \\[0.25em]
\hline
$\left\|e_{res}\right\|_{L^\infty}$  &9.78E-2	   &4.95E-2	        &2.47E-2	        &1.24E-2      \\
$\left\|e_{c}\right\|_{L^\infty}$                           &1.10E-3       &5.74E-4         &2.93E-4            &1.48E-4\\
$n_{tol}$                      &32              &55               &80                &106
\end{tabular*}
{\rule{\temptablewidth}{0.75pt}}
\end{center}
\end{table}

Based on the numerical results, we can see the iterative algorithm converges to the vortex with the prescribed $w$ and $m$, and the approximation results are very accurate. With closer initial guess, the iteration converges faster. The larger the spin index $m$ is, the bigger the core of the vortex is, which is consistent with results in \cite{Pego}. Comparing to results from the two examples, the vortices from the nonlinear optics in the second example have larger supports than those from the first example in rotating BEC, since there is no confining potentials in the NLS.

\subsection{Solving the modulation equations}Now, we can solve the modulation equations (\ref{me}) by some numerical discretization. 
Choose the time step size $\tau=\Delta t>0$ and denote the time steps by
$t_n:=n\tau, \, n=0,1,\ldots.$ To present the discretization, we denote the approximation of the involved quantities at $t_n$ as
\begin{align*}
&\binom{w^n}{\gamma^n}\approx \binom{w(t_n)}{\gamma(t_n)},\quad  A^n\approx A(t_n),\quad  G^n\approx G(t_n),\quad  g^n(\bx)\approx g(t_n,\bx),\\
& R^n(\mathbf{x})\approx R(t_n,\mathbf{x}),\quad \phi_{w}^n(\bx)\approx\phi_{w(t_n)}(\bx),\quad
\partial_w\phi_{w}^n(\bx)\approx\partial_w\phi_{w(t_n)}(\bx),
\end{align*}
and introduce the usual central finite difference operator for some grid function $f^n$ as
$$\delta_tf^n:=\frac{f^{n+1}-f^{n-1}}{2\tau},\qquad n=1,2,\ldots.$$
Then we discretize the modulation equations (\ref{me}) in time by a semi-implicit finite difference method as,
\begin{subequations}\label{me semi-dis}
\begin{align}
&A^n\delta_t\binom{w^n}{\gamma^n}=G^n,\quad n=1,2,\ldots,\label{me semi-dis: sigma}\\
&\delta_tR^n(\bx)=\frac{i}{2}\Delta\left[R^{n+1}(\bx)+R^{n-1}(\bx)\right]-iV(\bx)R^n(\bx)
+i(w^n-\delta_t\gamma^n)R^n(\bx)-g^n(\bx)\nonumber\\
&\qquad\qquad\ -i\delta_t\gamma^n\phi_w^n-\delta_t w^n\pa_w\phi_w^n,\quad \bx\in\RR^2,\ \ n=1,2,\ldots,\label{me semi-dis: R}
\end{align}
\end{subequations}
with initial values
$$w^0=w_0,\quad\gamma^0=\gamma_0,\quad R^0(\bx)=R_0(\bx),$$
and starting values at $t_1$ chosen as
\begin{align*}
&\binom{w^1}{\gamma^1}=\binom{w^0}{\gamma^0}+\tau\binom{w_t^0}{\gamma_t^0} ,\qquad \binom{w_t^0}{\gamma_t^0}=(A^0)^{-1}G^0,\\
&R^1(\bx)=R^0(\bx)+\tau\big[i\Delta R^{0}(\bx)-iV(\bx)R^0(\bx)
+i(w^0-\gamma_t^0)R^0(\bx)-g^0(\bx)\nonumber\\
&\qquad\quad\ -i\gamma_t^0\phi_w^0-w_t^0\pa_w\phi_w^0\big].
\end{align*}
For approximations in space, we again first truncate the whole space $\RR^2$ to the finite domain $I$ with periodic boundary condition and then use the Fourier pseudospectral discretization for the involved spatial derivatives and inner products. At every time step when $w$ is updated, corresponding $\phi_w,\partial_w\phi_w,\partial_w^2\phi_w$ are obtained by the algorithms described before.

The numerical scheme (\ref{me semi-dis}) gives second order accuracy in time and spectral accuracy in space, and is efficient thanks to the fast Fourier transform.  With the numerical approximations obtained from (\ref{me semi-dis}), the full solution $u$ to the NLS (\ref{nls}) could be recovered based on (\ref{ansatz}) with a trapezoidal rule towards the temporal integration as, 
$$u^n(\bx)=\exp\left\{-i\tau\left(\frac{1}{2}w^0+\frac{1}{2}w^n+\sum_{j=1}^{n-1}w^j\right)+i\gamma^n\right\}(\phi_{w^n}(\bx)
+R^n(\bx)),\quad n\geq1.$$
 With the full solution $u$ to the NLS, we are able to carry out the following test to justify the correctness of the modulation equations approach and the numerical discretizations. 
 
 \begin{table}[t!]
  \caption{Convergence on numerical solution of the modulation equations to NLS.}\label{tab:error}
  \vspace*{-10pt}
\begin{center}
\def\temptablewidth{1\textwidth}
{\rule{\temptablewidth}{0.75pt}}
\begin{tabular*}{\temptablewidth}{@{\extracolsep{\fill}}llllll}
                                 & $\tau=0.2$       &  $\tau/2$       & $\tau/4$           & $\tau/8$ \\[0.25em]\hline
$\left\|e_{u}\right\|_{L^\infty}$  &1.08E-1	   &3.80E-2	        &1.14E-2	        &4.10E-3      \\
\end{tabular*}
{\rule{\temptablewidth}{0.75pt}}
\end{center}
\end{table}

\begin{figure}[t!]
{$$
\begin{array}{cc}
\psfig{figure=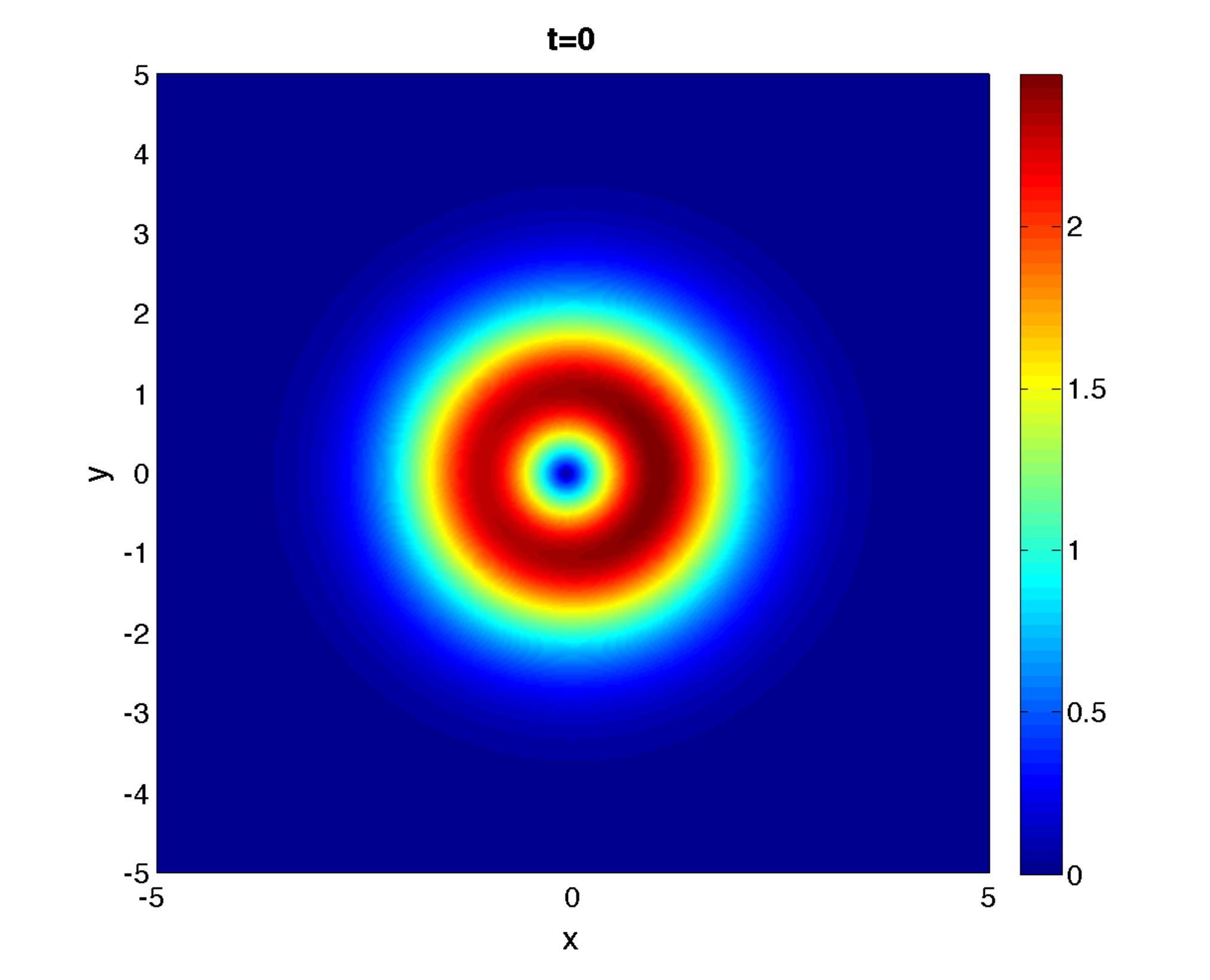,height=4cm,width=5.2cm}&\psfig{figure=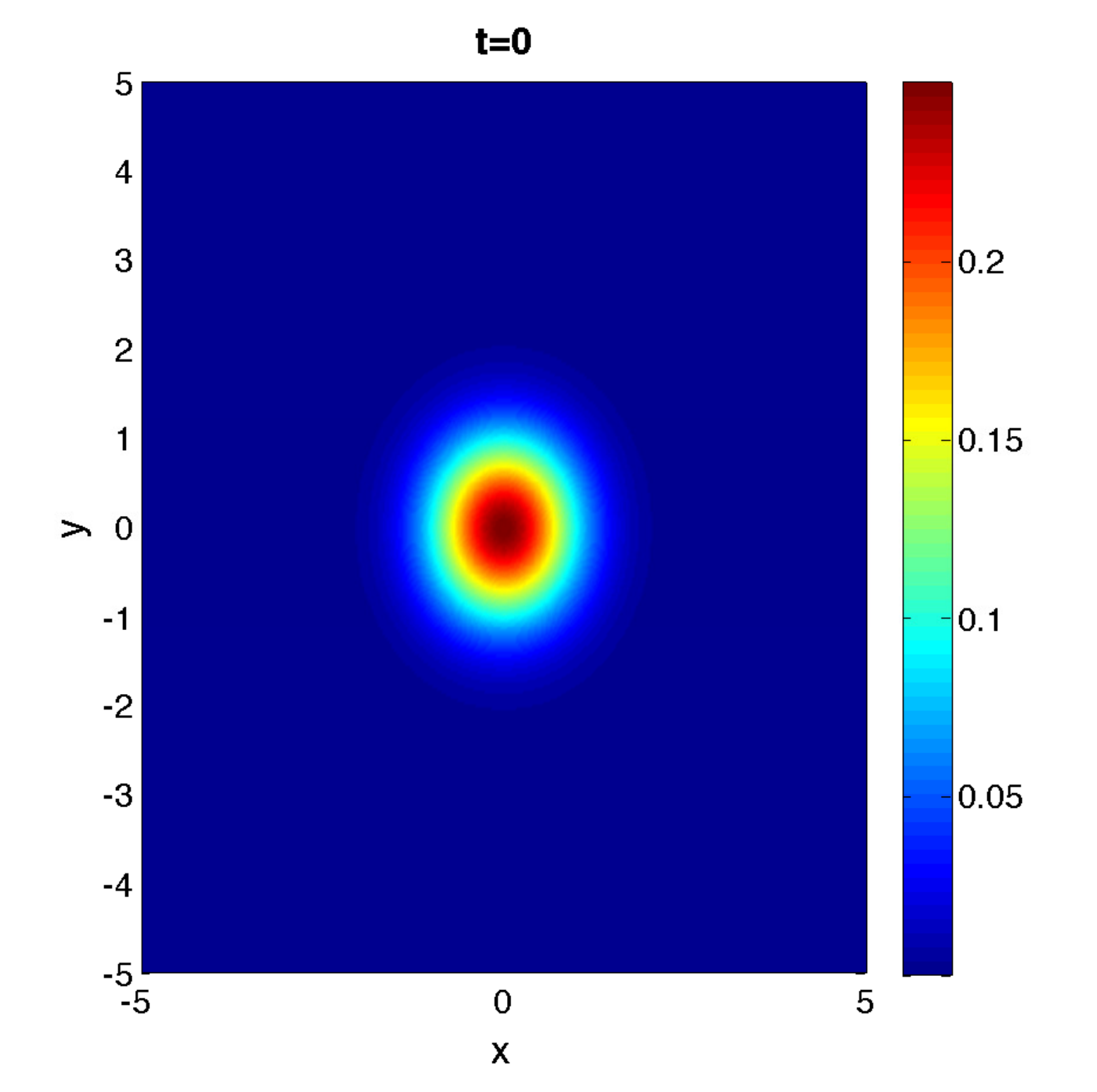,height=4cm,width=5.2cm}\\
\psfig{figure=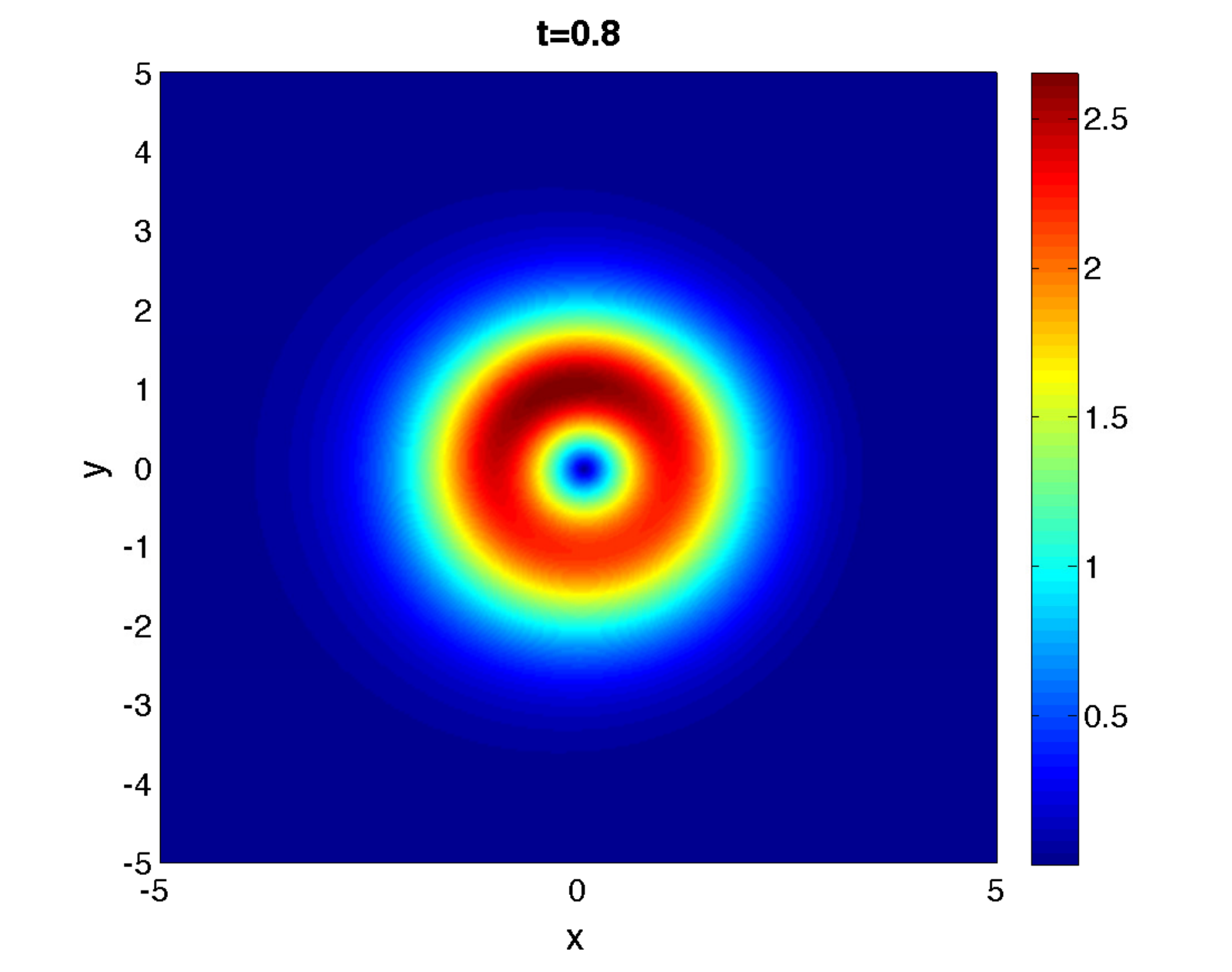,height=4cm,width=5.2cm}&\psfig{figure=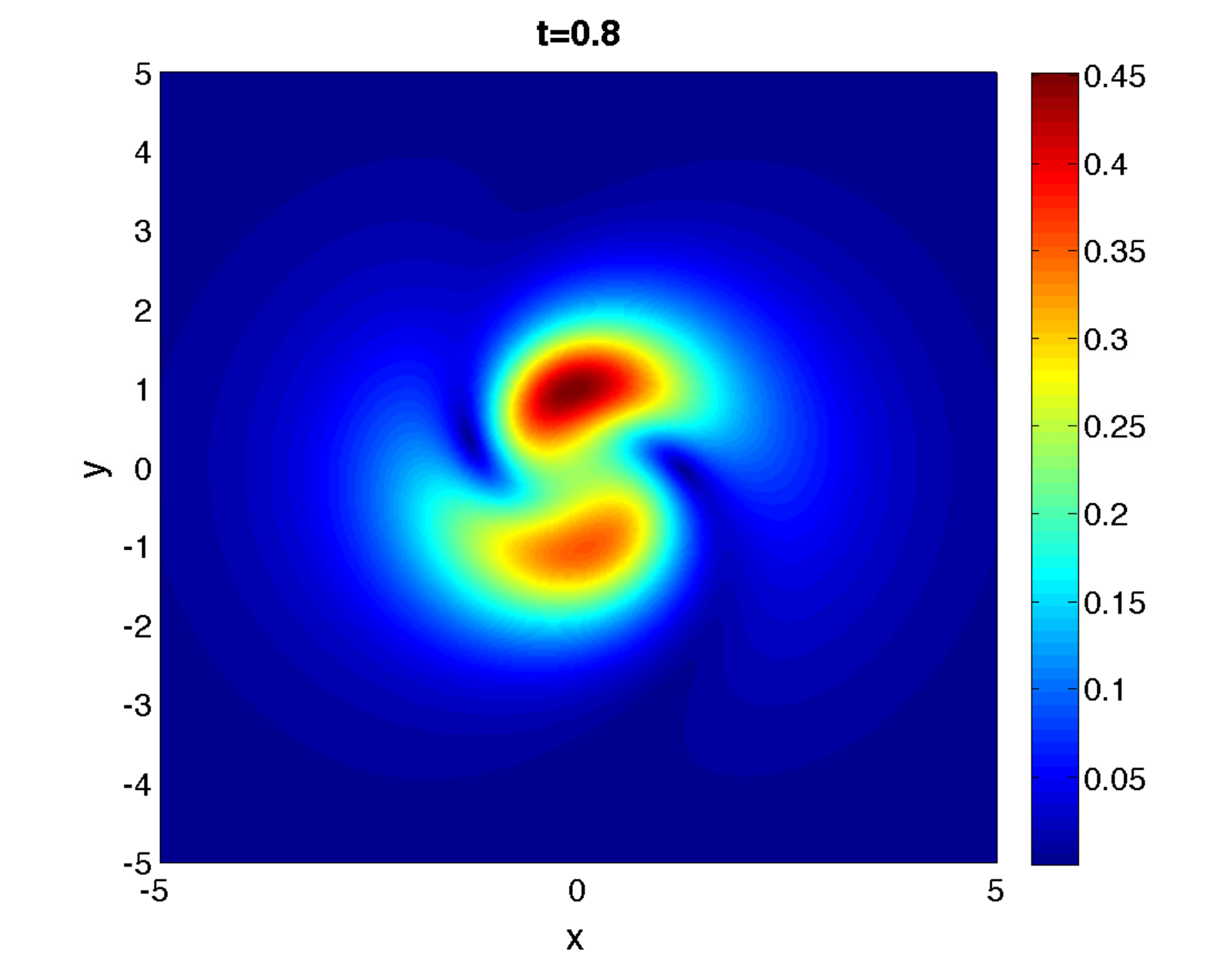,height=4cm,width=5.2cm}
\end{array}
$$
}
\caption{Contour plot of $|u(t,x,y)$ (left panel) and $|R(t,x,y)|$ (right panel) at $t=0.8$. }\label{fig:me}
\end{figure}
 
 We take the nonlinearity and the initial vortex as used in (\ref{vor:eg1}), i.e.
 $$\beta(\rho)=-0.5\rho,\ \ \rho\in\RR,\quad w_0=1.1,\quad m=1,\quad \gamma_0=1,$$
and the initial radiation prepared as
\begin{equation}\label{R0}
R_0(\bx)=\chi(\bx)-\frac{<\chi,\phi_w>}{\|\phi_w\|_{L^2}^2}\phi_w
 -\frac{<\chi,i\pa_w\phi_w>}{\|\pa_w\phi_w\|_{L^2}^2}i\pa_w\phi_w,\quad \chi(\bx)=\frac{\fe^{-|\bx|^2}}{4},
 \end{equation}
in order to satisfy the orthogonality condition (\ref{oc}) at $t=0$. We then solve the NLS (\ref{nls}) by the proposed method on the computational domain $I=[-8,8]\times[-8,8]$ and compare with the solution obtained from the directly solving (\ref{nls}), saying by the time-splitting spectral method \cite{BaoCai}. The error 
$$e_u(\bx):=u(t_n,\bx)-u^n(\bx),$$
at $t=t_n=0.8$ is given in Table \ref{tab:error} under different step size $\tau$ (with spatially fine mesh). The profiles of the solutions $u$ and $R$ are given in Figure \ref{fig:me}.

From the numerical results in Table \ref{tab:error} and Figure \ref{fig:me}, we see the modulation equations approach solves the NLS correctly and accurately. The dynamics of the vortex and radiation are captured separately. The radiation wave is trapped by the confining potential and its amplitude does not decrease.

\begin{figure}[t!]
{$$
\begin{array}{cc}
\psfig{figure=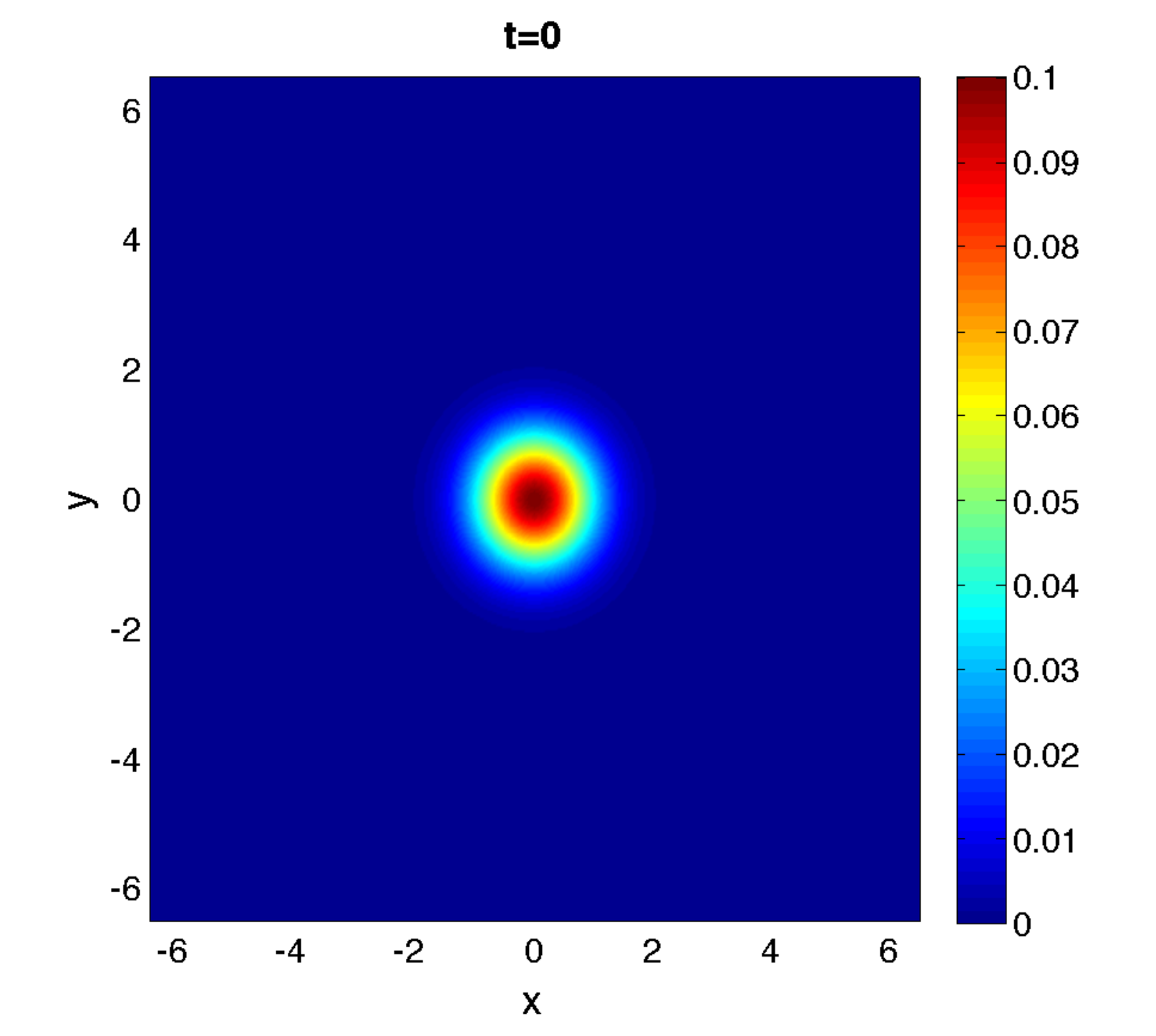,height=4.3cm,width=5.3cm}&\psfig{figure=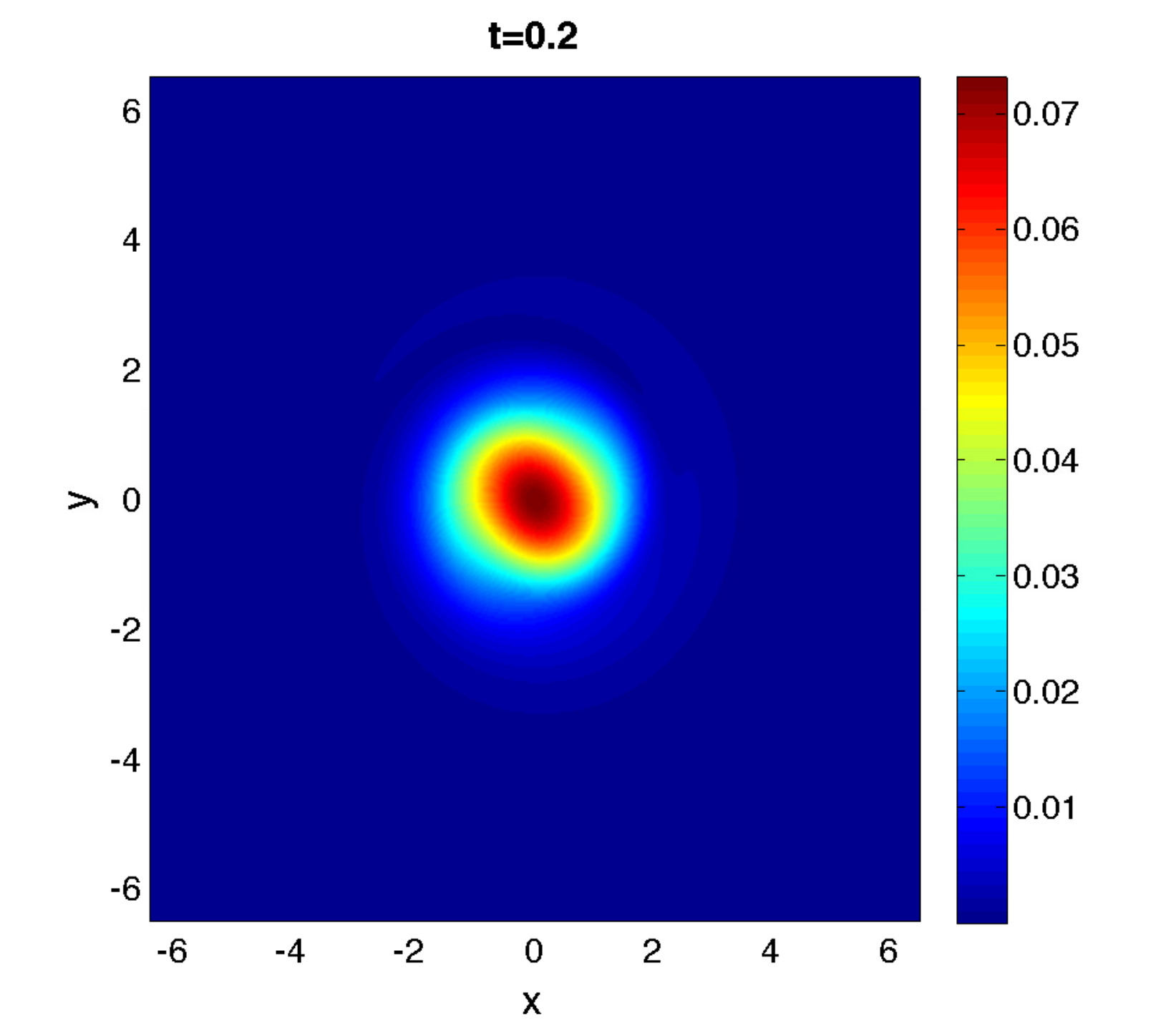,height=4.3cm,width=5.3cm}\\
\psfig{figure=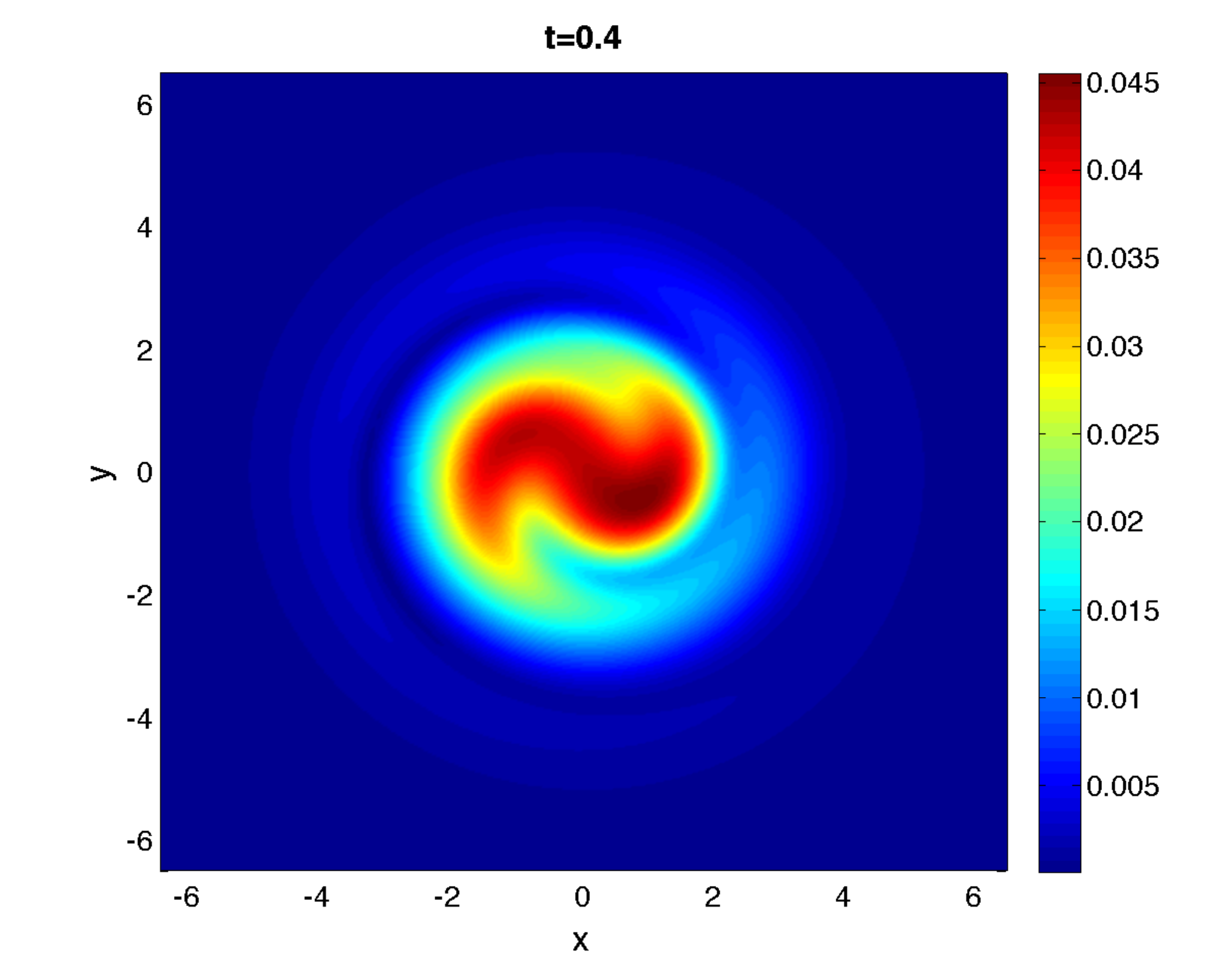,height=4.3cm,width=5.3cm}&\psfig{figure=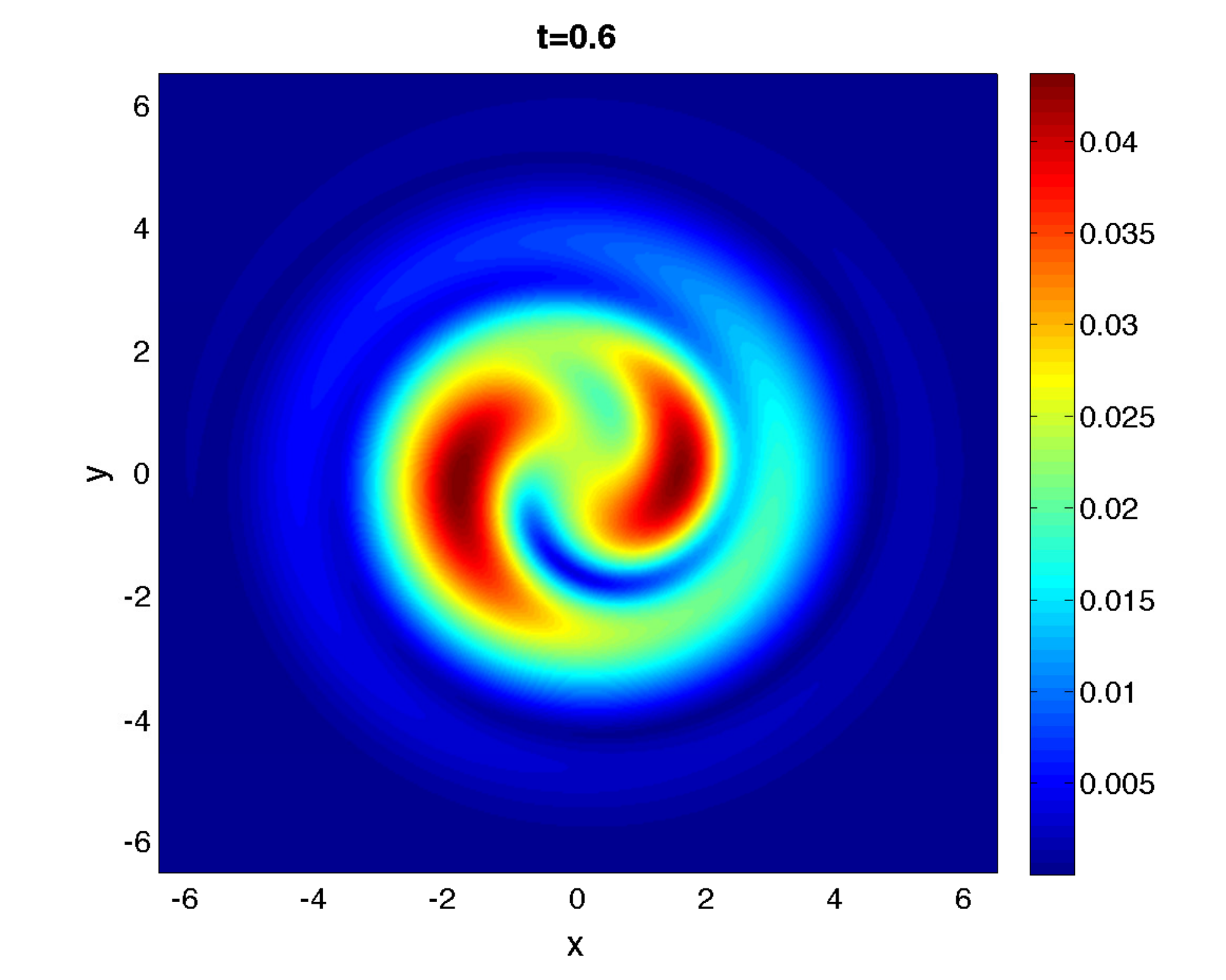,height=4.3cm,width=5.3cm}
\end{array}
$$
}
\caption{Contour plot of $|R(t,x,y)|$ at different $t$. }\label{fig:me2}
\end{figure}

To see the nonlinear scattering of free (from external potential) radiation, we consider the setup in example (\ref{vor:eg2}), i.e.
$$V(\bx)=0,\quad \lambda=-2,\quad w_0=-0.5,\quad \gamma_0=1,$$
 and solve the dynamics of a vortex of index $m=1$ and radiation. Choosing the initial radiation similarly as (\ref{R0}) with $\chi(\bx)=0.1\fe^{-|\bx|^2}$, the dynamics of $R(t,\bx)$ are shown in Figure \ref{fig:me2}. The results show that the initial bump gradually expand to far field and its amplitude decrease to zero. The whole system will turn to a steady state \cite{Soffer1,Zhao1,Zhao2}. A very interesting observation from both Figure \ref{fig:me} and Figure \ref{fig:me2} is that the radiation starts to rotate as time evolves. 

\begin{figure}[t!]
{$$
\begin{array}{ccc}
\psfig{figure=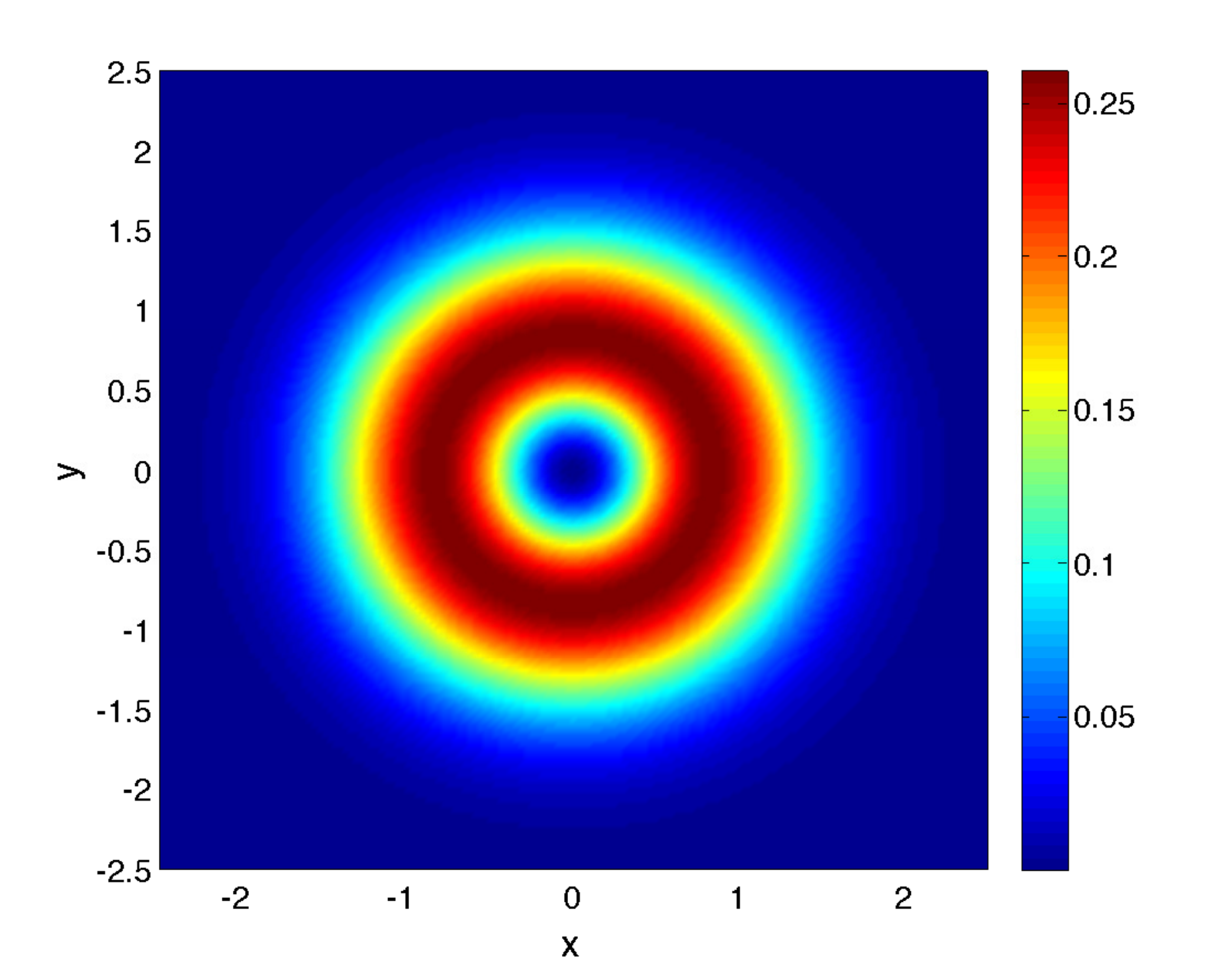,height=4cm,width=4cm}&\psfig{figure=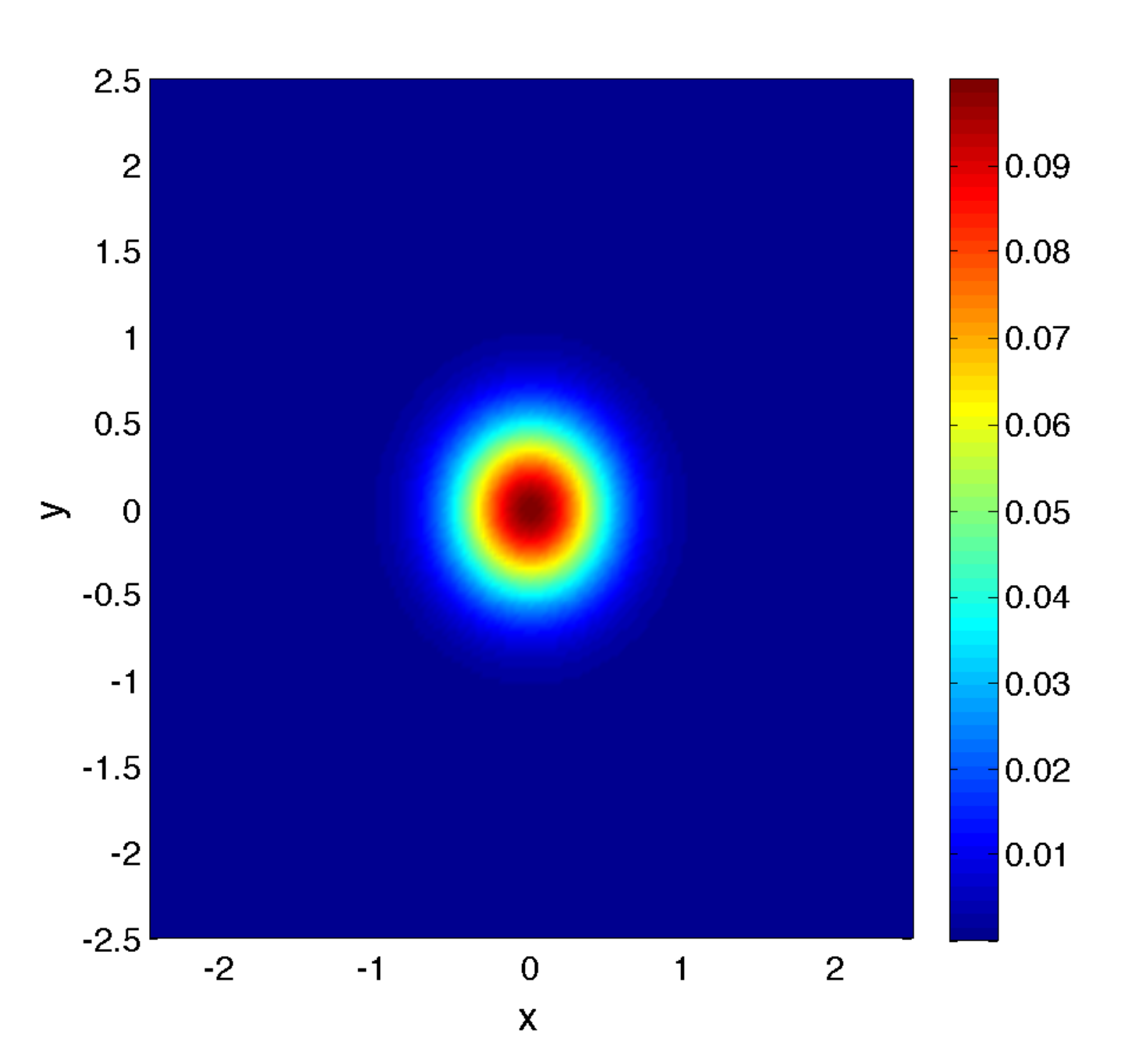,height=4cm,width=4cm}&
\psfig{figure=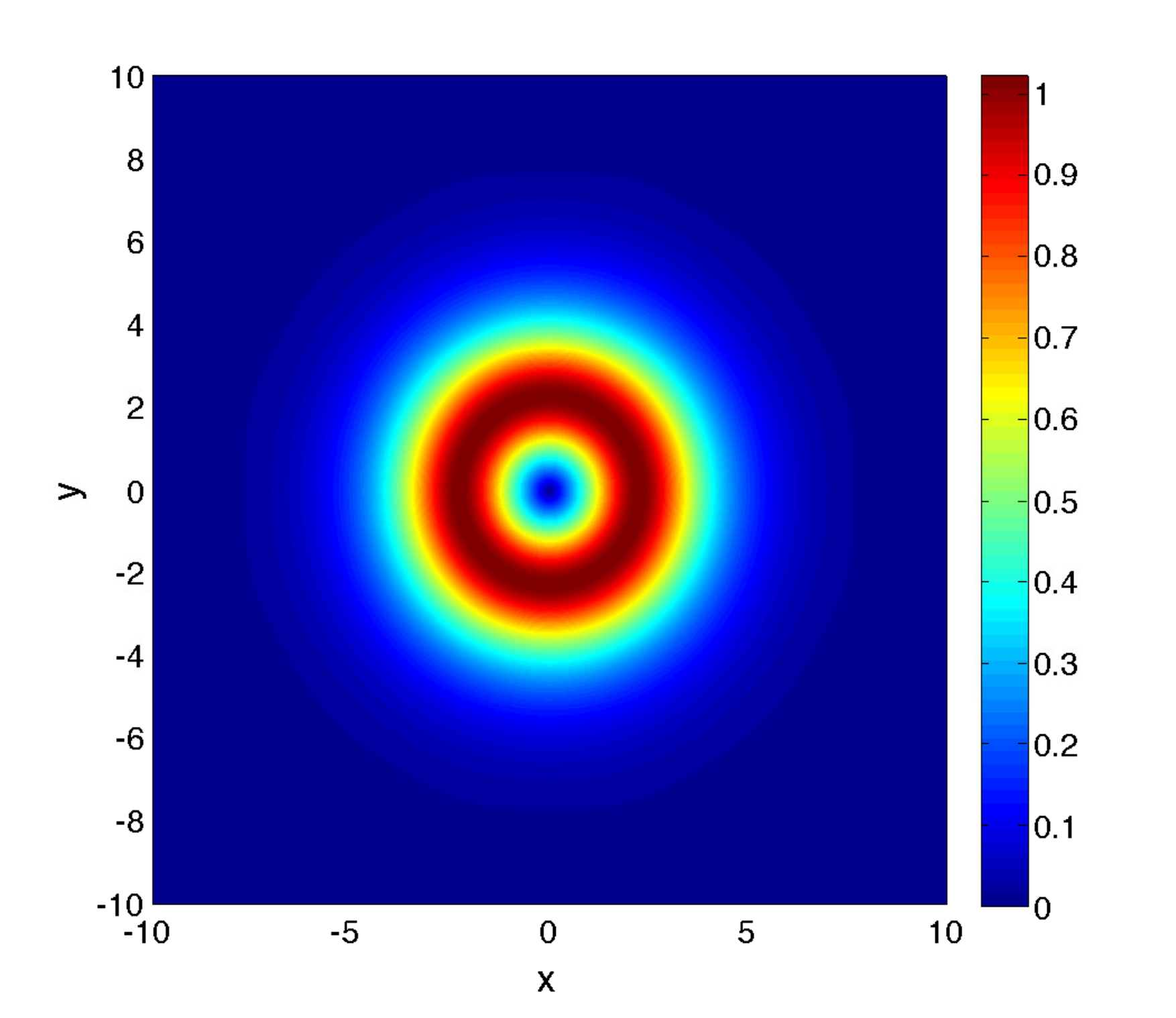,height=4cm,width=4cm}
\end{array}
$$
}
\caption{Trapping potential $V(x,y)$ (left), initial radiation $|R(0,x,y)|$ (middle) and vortex $|\phi_{w(0)}(x,y)|$ (right). }\label{fig:tunnel}
\end{figure}

\begin{figure}[t!]
{$$
\begin{array}{ccc}
\psfig{figure=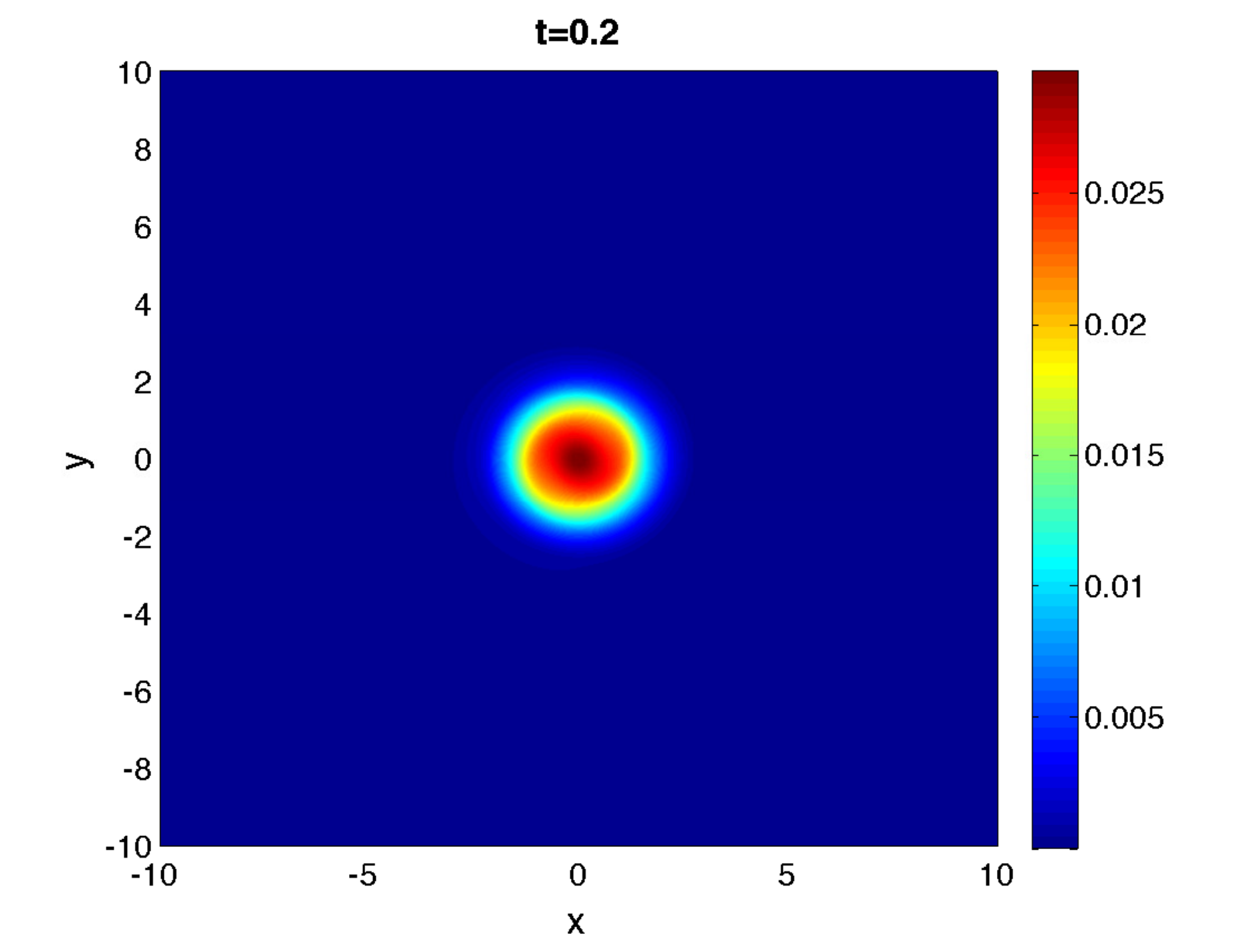,height=4cm,width=4cm}&\psfig{figure=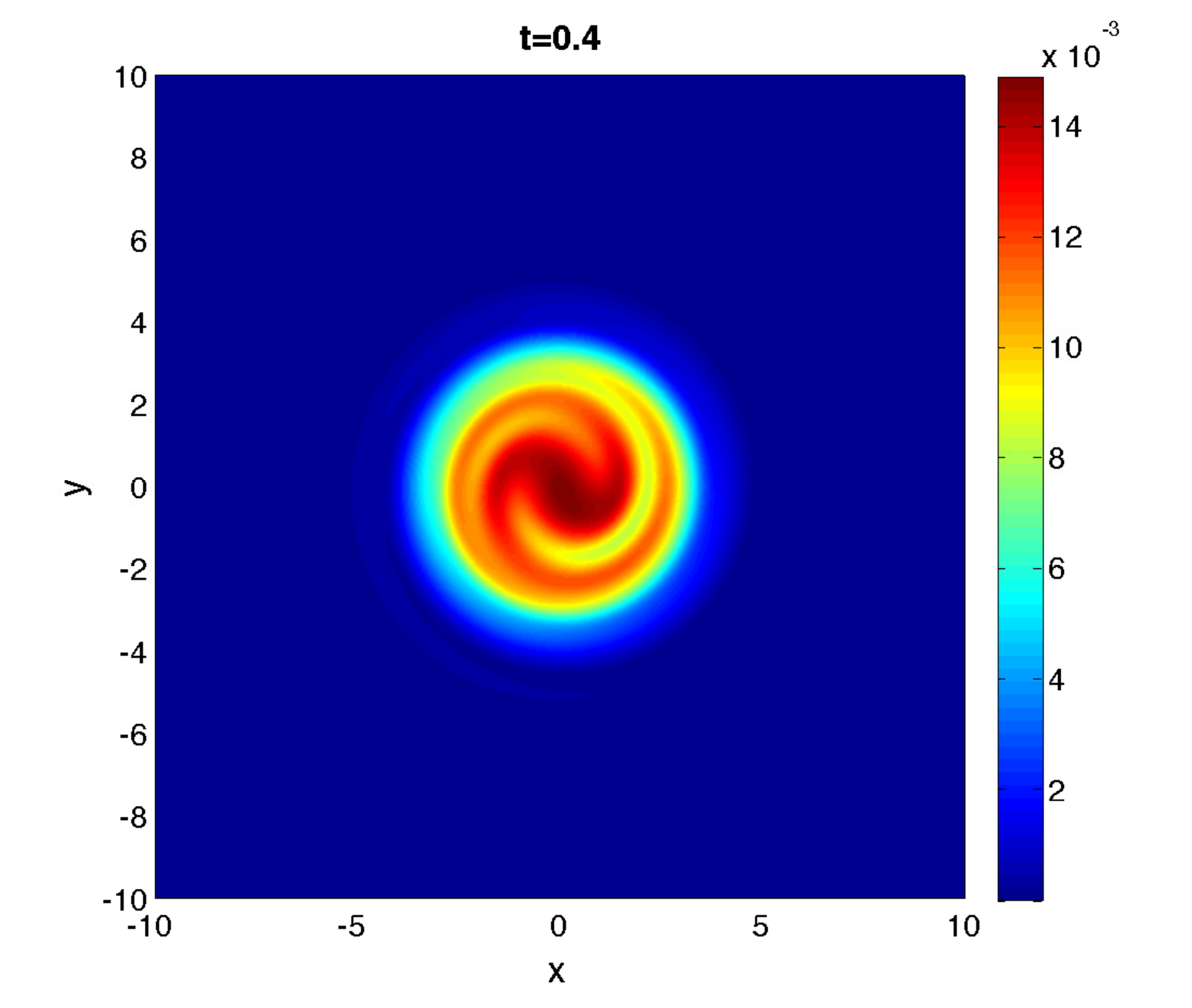,height=4cm,width=4cm}&\psfig{figure=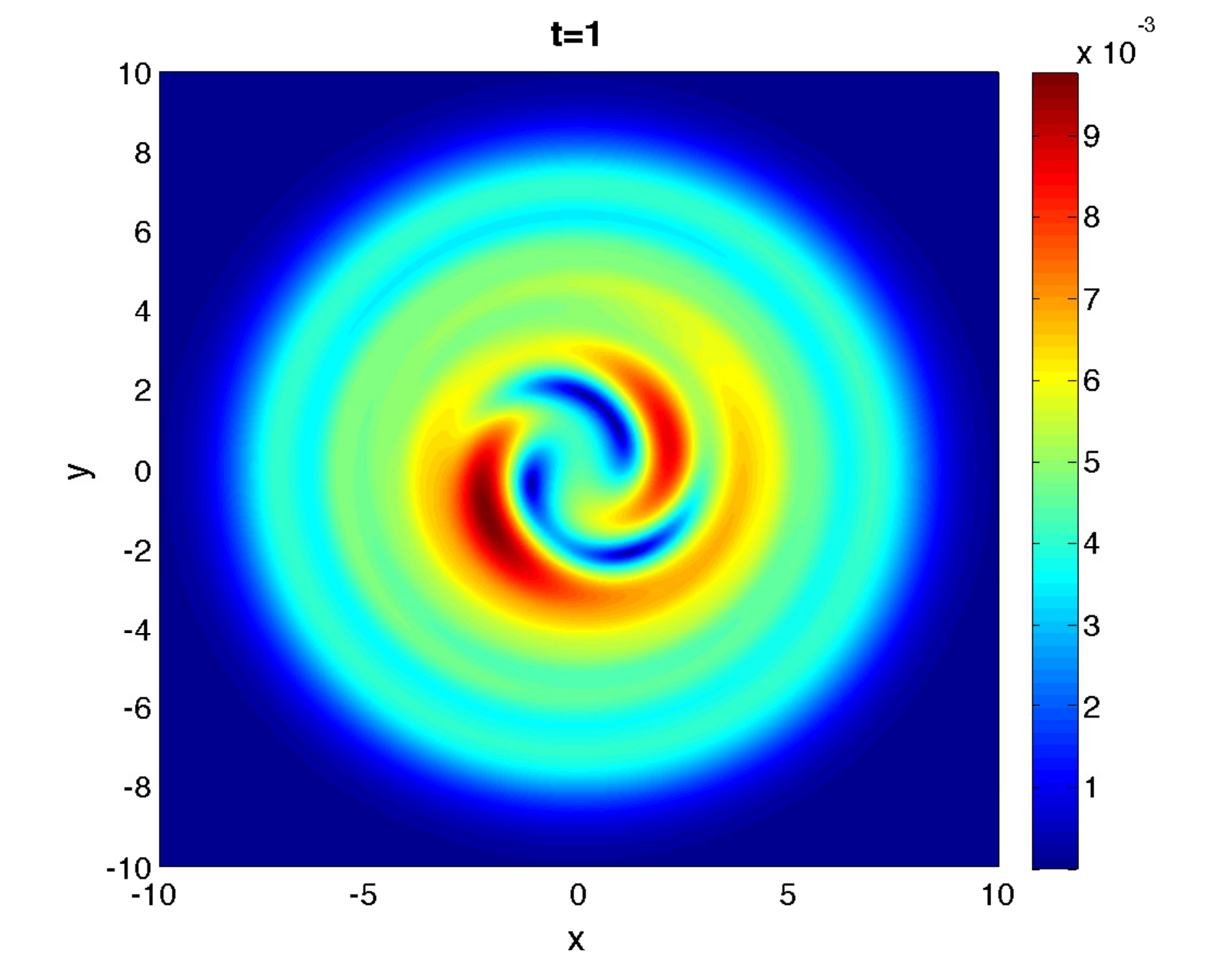,height=4cm,width=4cm}
\end{array}
$$
}
\caption{Tunnelling of radiation wave through potential: $|R(t,x,y)|$ at different $t$. }\label{fig:me3}
\end{figure}

At last but not least, we consider the quantum tunnelling \cite{Fleurov,Razavy} of vortex and radiation wave:
$$\lambda=-1.5,\quad w_0=-0.5,\quad \gamma_0=1,\quad \chi(\bx)=0.1\fe^{-4|\bx|^2},$$
 living in a trapping potential, i.e.
$$V(\bx)=|\bx|^2\fe^{-\sqrt{2}|\bx|^2}.$$
The profile of the potential $V(\bx)$, initial radiation $R(0,\bx)$ and vortex $\phi_{w(0)}(\bx)$ are given in Figure \ref{fig:tunnel}.
The dynamics of $R(t,\bx)$ are shown in Figure \ref{fig:me3}. From the results, we can see the radiation rotates and disperses through the trapping potential. It becomes bumpy after passing through the potential wall. The ring shaped bumps of the tunnelled waves seems to be the two dimensional analog of the blip phenomena as described in one dimension in \cite{Dekel,Dekel1,Dekel2}. As shown in \cite{Cohen}, the bumps will turn into jets if the potential is anisotropic. However, when the trapping potential is anisotropic, the vortex state in (\ref{inls}) would become non-symmetric, and formulation (\ref{form}) or (\ref{rho}) becomes invalid. More efforts are needed for the algorithm and simulations, which would be our future investigation.

\section{Conclusion}\label{sec:conc}
We solved the multichannel solution of a vortex and radiation in the nonlinear Schr\"{o}dinger equation (NLS) by means of the modulation equations approach. We firstly derived the fully modulation equations for dynamics of the vortex and radiation, and then solved the modulation equations via numerical methods, where an accurate iterative algorithm has been given to obtain vortices with prescribed energy and spin index. Numerical tests and simulations of nonlinear scattering were given. The approach and strategy could be generalised in future to solve the NLS with anisotropic potential and be applied to study the jetlike tunnelling effect in \cite{Cohen}. 

\section*{Acknowledgements}
This work was partially supported by a grant from the Simons Foundation (\#395767 to Avraham Soffer).
A. Soffer is partially supported by NSF grant DMS 1201394. X. Zhao is supported by the French ANR project MOONRISE ANR-14-CE23-0007-01.
Part of this work was done when the authors were visiting the School of Mathematics and Statistics, Central China Normal University, January, 2016.

\end{document}